\documentclass{aa}

\usepackage{graphicx}
\usepackage{txfonts}
\usepackage{natbib}

\usepackage{amsfonts}
\usepackage{amssymb}

\begin{document}

\title{Excitation of ion-acoustic waves by \\ non-linear finite-amplitude standing Alfv\'en waves}
\author{Jaume Terradas\inst{1}, Adolfo F.- Vi\~{n}as\inst{2,3}, Jaime A. Araneda\inst{4}}

\institute{$^1$Departament de F\'\i sica, Universitat de les Illes Balears (UIB),
E-07122, Spain \\   Institute of Applied Computing \& Community Code (IAC$^3$),
UIB, Spain\\ \email{jaume.terradas@uib.es}\\
$^2$Department of Physics \& the Institute for Astrophysics and Computational Sciences (IACS), Catholic University of America, Washington-DC, 20064, U.S.A \\
$^3$NASA Goddard Space Flight Center, Heliospheric Science Division, Geospace Physics Laboratory, Mail Code 673, Greenbelt, MD 20771, U.S.A. \\ \email{adolfo.vinas@nasa.gov}\\
$^4$Departamento de F\'\i sica, Facultad de Ciencias F\'\i sicas y Matem\'aticas, Universidad de Concepci\'on, Concepci\'on, Chile\\ \email{jaraneda@udec.cl}\\
}

\abstract{
We investigate, using a multi-fluid approach, the main properties of standing  ion-acoustic modes driven by nonlinear standing Alfv\'en waves. The standing character of the Alfv\'enic pump is because we study the superposition of two identical circularly polarised counter-propagating waves. We consider parallel propagation along the constant magnetic field and we find that left and right-handed modes generate via ponderomotive forces the second harmonic of standing ion-acoustic waves. We demonstrate that parametric instabilities are not relevant in the present problem and the secondary ion-acoustic waves attenuate by Landau damping in the absence of any other dissipative process. Kinetic effects are included in our model where ions are considered as particles and electrons as a massless fluid, and hybrid simulations are used to complement the theoretical results. Analytical expressions are obtained for the time evolution of the different physical variables in the absence of Landau damping. From the hybrid simulations we find that the attenuation of the generated ion-acoustic waves follows the theoretical predictions even under the presence of a driver Alfv\'enic pump. Due to the nonlinear induced ion-acoustic waves the system develops density cavities and an electric field parallel to the magnetic field. Theoretical expressions for this density and electric field fluctuations are derived. The implications of these results in the context of standing slow mode oscillations in coronal loops is discussed.}

   \titlerunning{Nonlinear excitation of standing  ion-acoustic waves}

   \maketitle

	\section{Introduction}

	Alfv\'en waves are thought to play an important role in coronal heating, solar wind acceleration and the development of turbulence in solar wind plasmas. The reader is referred to the works of \citet{hasegawauberoi1982,cross1988,cramer2001} for a broad introduction on the topic of Alfv\'en waves. Circularly polarised Alfv\'en waves are an exact solution to the MHD equations even when the amplitude of the wave is large.  The fluid approach has been used in the past in the description of Alfv\'en waves for reasons of mathematical tractability, however fluid theory  predicts a dramatic dependence of the occurrence of nonlinear processes on the value of the plasma-$\beta$ \citep[e.g.][]{wonggolstein1986}. For this reason kinetic theory throughout the Vlasov equation provides the most general description of a plasma \citep[see][for applications to nonlinear Alfv\'en waves]{spangler1989}.
	
	The analysis of Alfv\'en waves over the last decades has been mostly focused on propagating waves because of their academic interest and also because of their practical applications to both laboratory and space plasmas. \citet{hollweg71} showed that linearly polarised Alfv\'en waves propagating parallel to the magnetic field are able to drive density enhancements due to gradients in the wave magnetic-field pressure. Contrastingly, circularly polarised low-frequency Alfv\'en wave propagating along a constant magnetic field, are known to be parametrically unstable \citep[e.g.][]{derby1978,goldstein1978}, a process that  has received significant attention in the scientific community. It can be understood as a wave-wave interaction mechanism in which a large-amplitude wave or pump couples nonlinearly with smaller-amplitude fluctuations according to the laws of energy and momentum conservation. Three types of parametric instabilities are found, modulational, beat and decay \citep[e.g.][]{hollweg1994}. The decay instability is of particular interest in the solar wind since it might provide the backward propagating Alfv\'en waves that are thought to be necessary for the development of magnetohydrodynamic turbulence. The analysis of these instabilities has been mostly carried out in the single fluid approach and dispersive effects arising from the cyclotron motion of single-ion species have been also included in the analysis. Moreover, kinetic effects have been also considered in the problem of parametric instability by \citet{araneda1998, aranedaetal2007,aranedaetal2008}. These authors, using linear Vlasov theory and hybrid simulations found that kinetic effects, such us Landau damping (LD hereafter), break the degeneracy of the mode-coupling solutions of fluid theory and are important even for very low proton plasma-$\beta$.
	
	On the contrary, standing Alfv\'en waves have received much less attention in the literature. They can be formed at any place where trapped or reflected Alfv\'en waves propagate in opposite directions. They can be viewed as the superposition of two identical counter-propagating waves. The most clear examples of standing Alfv\'en waves are found in the magnetospheres of planets such as the Earth, Jupiter and Mercury \citep[see for example][]{cummingsetal1969,manneretal2018,khurana1993}. Other examples are related to standing Alfv\'enic transverse loop oscillations often reported in the solar atmosphere due to energy releases produced by flares \citep[e.g.][]{nakaetal99,schrijetal02,aschetal02}. A nearby perturbation is able to impact laterally coronal loops and excite Alfv\'en like waves travelling along these closed magnetic flux tubes. These waves are almost fully reflected back at the solar photosphere where the density is several orders of magnitude larger than that in the solar corona \citep[see the recent review of][]{nakaetal2021}. Laboratory plasmas is another example where the presence of an initial Alfv\'enic pump can generate standing Alfv\'en waves. For example, \citet{danielsonetal2002} studied the generation and damping of standing ion-acoustic waves in Penning-Malmberg electron traps.
	
	Using a single fluid approach \citet{chianoliverira1994,chianoliverira1996,oliverirachian1996} studied the effects of parametric instabilities on standing waves with applications to the magnetosphere. These authors developed a theory of magnetohydrodynamic (MHD) parametric instabilities driven by a left-hand or right-hand circularly polarised standing Alfv\'en wave in a low plasma-$\beta$ under very specific initial conditions. \citet{isofman2004,isofman2011} using both the MHD approach and hybrid simulations investigated numerically the generation of an electric field parallel to the magnetic field, due to non-linear induced ion-acoustic waves produced by the presence of a  standing Alfv\'enic pump. In a different context, \citet{mottez2015J} also explored the generation of a parallel electric field using PIC (particle in cell) simulations when two counter-propagating Alfv\'en waves are excited. However, the mechanism proposed by this author is not related to the ponderomotive force that it is usually operating under such conditions. The ponderomotive force,  see the review of \citet{lundinguglielmi2006}, generates density enhancements  periodically in time and space which are normally called cavities and depend on gradients in density or electric fields. The periodicity depends on the value of gas pressure of the system and the amplitude of these enhancements is related quadratically with the amplitude of the initial transverse excitation \citep[see also][]{rankinetal94,tikhoetal1995}. Interestingly, ponderomotive forces can lead to separation of species, although this topic is not addressed in the present work since the quasi-neutrality condition is imposed for the plasma composed of kinetic protons and the treatment of electrons is that of a fluid. Recently, \citet{davidetal2018} using the multi-fluid approach have modelled high-frequency waves in partially ionised plasmas akin to those in quiescent solar prominences. These authors have shown that nonlinear Alfv\'en waves generate density and pressure perturbations and that the friction due to collisions dissipates a fraction of the wave energy, which is transformed into heat raising the temperature of the plasma.

The main goal of the present paper is to understand, using the multi-fluid approach and hybrid simulations, the generation of ion-acoustic modes driven nonlinearly by standing finite amplitude transverse waves in 1D. We show both analytically and numerically, that in our set-up there are no parametric instabilities and that the ponderomotive force is the main responsible for the dynamics of the system. Our analysis includes kinetic effects and addresses the relevant question of how Landau damping of ion-acoustic waves due to wave-particle interactions is affected by the presence of the Alfv\'enic driver. We further propose some simple applications of the results presented in this paper.

\section{Basic plasma model configuration}\label{sectconf}

We assume a quasi-neutral electron-proton plasma. The background magnetic field, $\bf B_0$, is constant and pointing in the $x-$direction. The equations are solved in this direction only although the velocity, electric and magnetic components in the $y$ and $z$ directions are retained in the description, i.e., we concentrate on purely parallel wave propagation to the magnetic field. We adopt the multi-fluid approach and use the standard massless fluid-electron model. Nevertheless, protons and electrons are allowed to have different temperatures ($T_{\rm p}$ and $T_{\rm e}$) and this leads to the presence of LD in the system, especially in the situation when $T_{\rm p} \ge T_{\rm e}$.

Our interest is on standing waves and therefore periodic boundary conditions are applied at the edge of the spatial domain, where nodes in the three components of the velocity are imposed at $t=0$ and maintained during the time evolution. This mimics the effect of line-tying. The domain is defined in the range $0\leq x \leq L$ where $L$ is the size of the system.

\section{Preliminaries: Ion-acoustic standing waves}

We start describing the results for standing ion-acoustic waves that are eigenmodes of the configuration and not forced oscillations due to a driver. Since the longitudinal magnetic field is assumed constant, with zero transverse magnetic fluctuations, it plays a passive role in this section. The results are well known for propagating waves but for standing waves there are some points that need to be explained in detail, being the most relevant that a standing ion-acoustic wave is attenuated by Landau damping. 
 
In the multi-fluid approach each species, denoted by the subindex, $s$, satisfy the continuity equation and the momentum equation,
\begin{align}\label{eqcont}
\frac{\partial n_s}{\partial t}+\frac{\partial (n_s {u_{xs}}) }{\partial x}&=0,\\
\label{eqmotion0}
\frac{\partial {\bf u_{s}}}{\partial t}+ { u_{xs}}\frac{\partial {\bf u_s}}{\partial x}&=\frac{q_{s}}{m_{s}}\left({\bf{E}} + \frac{{\bf u_{s}} \times {\bf B}}{c}\right)-\frac{1}{m_{s}n_{s}}\frac{\partial p_{s}}{\partial x} {\bf {\hat e}_x},
\end{align}
where collisions are neglected and the variables have their usual meaning. The mass density of each species is defined as $\rho_s=m_s n_s$. In the present case we consider protons and electrons only, but we assume that electrons are massless. This simplifies the problem  allowing us a direct comparison with the hybrid numerical simulations described later. The previous equations are completed with Maxwell equations and the closure of the system of equation in fluids is often done with the adiabatic gas law. These equations lead to the MHD equations when the single fluid approximation is used.

For a constant horizontal magnetic field the linearised longitudinal momentum equation reduces for the parallel velocity component to the following equation
\begin{eqnarray}\label{eq:vpareq00}
m_s\frac{\partial \delta u_{x s}}{\partial t}=q_s \delta E_x-\frac{1}{n_{0s}}\frac{\partial \delta p_s}{\partial x}. 
\end{eqnarray}
where $\delta u_{xs}$, $\delta E_x$ and $\delta p_s$ represent the parallel velocity, parallel electric field, and the pressure fluctuations respectively, while $n_{0s}$ is the equilibrium density. We eliminate the longitudinal component of the parallel electric field by assuming massless electrons, therefore the previous equation when applied to electrons simplifies to ($q_e=-e$)
\begin{eqnarray}\label{epar0}
\delta E_x=-\frac{1}{e n_{0}}\frac{\partial \delta p_e}{\partial x}, 
\end{eqnarray}
where we have neglected the term on the left hand side of Eq.~(\ref{eq:vpareq00}). Imposing quasi-neutrality means that $n_{\rm 0p}=n_{0e}\equiv n_0$, and for the same reason the proton and electron density fluctuations must be the same,  i.e., $\delta n_p=\delta n_e \equiv \delta n $. Equation~(\ref{epar0}) is then introduced in Eq.~(\ref{eq:vpareq00}) for protons. Using the standard adiabatic law for the perturbations we have
\begin{equation}\label{eqdensp}
\frac{\delta p_{s}}{p_{0s}}=\gamma_s \frac{\delta n_{s}}{n_{0s}}.
\end{equation}
These equations are complemented with the continuity equation which implies that  $\delta u_{x p}= \delta u_{x e}$ and therefore hereafter we refer to this variable as  $\delta u_{x}$.

The resulting system of equations is
\begin{align}\label{vpareqs0}
\frac{\partial \delta u_x}{\partial t}&=-\left(c^2_{\rm sp}+c^2_{\rm se}\right)\frac{1}{n_{0}}\frac{\partial \delta n}{\partial x},\\
\frac{\partial \delta n}{\partial t}&=-n_0 \frac{\partial \delta u_x}{\partial x}.\label{vpareqs01}
\end{align}
Where we have used the definition of the sound speeds, $c^2_{\rm se}=\gamma_{\rm e} \frac{k_{\rm B}}{m_{\rm p}} T_{\rm e}$ and $c^2_{\rm sp}=\gamma_{\rm p} \frac{k_{\rm B}}{m_{\rm p}} T_{\rm p}$, being $\gamma_{\rm e}$ and $\gamma_{\rm p}$ the corresponding values of the polytropic index for electrons and protons and $T_{\rm e}$ and $T_{\rm p}$ are the equilibrium temperatures of the two species.

We rewrite Eq.~(\ref{vpareqs0}) as a single equation for $\delta u_x$,
\begin{eqnarray}\label{vpareq02new}
\frac{\partial^2 {\delta u_x}}{\partial t^2}-C^2_{\rm s}\frac{\partial^2 {\delta u_x}}{\partial x^2}=0.
\end{eqnarray}
where we have introduced $C^2_{\rm s}={c^2_{\rm se}+c^2_{\rm sp}}$. This equation is solved applying initial and boundary conditions. Since we are interested on standing waves in a finite domain these conditions are
\begin{align}\label{vcond0}
{\delta u_x} (x,t=0)&=f(x),\\
\frac{\partial \delta u_x}{\partial t}(x,t=0)&=0,\\
{\delta {u_x}}(0,t)&={\delta u_x}(L,t)=0,
\end{align}
where $f(x)$ is the spatial profile of the initial perturbation.
The solution to the wave equation together with the previous initial and boundary conditions is known to be
\begin{eqnarray}\label{solvpar2vp0}
{\delta u_x}(x,t)=\sum^\infty_{m=1} c_m \cos \left({C_{s} \frac{m \pi} {L} t}\right) \sin\left({\frac{m \pi} {L} x}\right),
\end{eqnarray}
where the ${c_m}$ are the Fourier coefficients given by
\begin{eqnarray}\label{solvparcn}
c_m=\frac{2}{L}\int^L_0 f(x) \sin\left({\frac{m \pi} {L} x}\right)  dx, \mbox{ } m=1,\, 2,\, \ldots
\end{eqnarray}
\noindent Equation~(\ref{solvpar2vp0}) satisfies the expected boundary conditions (nodes) at $x=0$ and $x=L$ in velocity. It represents a standing solution and can be viewed as the superposition of two counterpropagating waves. The standing solution has zero propagation speed since the two counterpropagating waves have the same phase velocity but of opposite sign. 

The dispersion relation that we obtain is simply
\begin{eqnarray}\label{omegadis}
\omega=k\, C_{\rm s},
\end{eqnarray}
which corresponds to ion-acoustic waves and 
\begin{eqnarray}\label{kdis}
k=\frac{m \pi}{L}, \mbox{ } m=1,\, 2,\, \ldots
\end{eqnarray}
is the discrete wavevector that matches the boundary conditions.

For the density perturbation we also obtain an homogeneous wave equation whose solution is calculated using Eq.~(\ref{solvpar2vp0}) and Eq.~(\ref{vpareqs01}),
\begin{eqnarray}\label{solvpar2nm}
\delta n(x,t)=-\frac{n_0}{C_{\rm s}} \sum^\infty_{m=1} c_m \sin \left({C_{\rm s} \frac{m \pi} {L} t}\right) \cos\left({\frac{m \pi} {L} x}\right).
\end{eqnarray}
Hence,  density perturbations are $90^\circ$ out of phase in time and space with respect to velocity. The density fluctuation, contrary to the parallel velocity, has antinodes at $x=0$ and $x=L$.  These results, although simple, are useful to understand the situation when the Alfv\'enic driver is present.   

The expression for the electric field is obtained from Eq.~(\ref{epar0}) using Eq.~(\ref{solvpar2nm}) and reads
\begin{eqnarray}\label{epar0ex}
\delta E_x (x,t)=-\frac{B_0}{c}\frac{1}{\omega_{\rm cp}}\frac{c^2_{\rm se}}{C_{\rm s}} \sum^\infty_{m=1} \left(\frac{m \pi} {L}\right) c_m \sin \left({C_{\rm s} \frac{m \pi} {L} t}\right) \sin\left({\frac{m \pi} {L} x}\right).
\end{eqnarray}
where the equilibrium magnetic field, $B_0$, and $\omega_{\rm cp}$ (the cyclotron frequency for protons) appear for normalisation purposes only.
    
\subsection{Kinetic effects on ion-acoustic standing waves}\label{kinetion}

The aim of this section is to introduce kinetic effects by implementing a relatively simple method of \citet{araneda1998,aranedaetal2007} to incorporate Landau damping in the multi-fluid system of equations, previously described. The procedure considers the closure between pressure and density moments relationship by a fully kinetic treatment of the ion-acoustic wave dispersion analysis via the Vlasov equation, instead of using the traditional polytropic closure of the fluid equations. This avoids the inadequacies of fluid models in describing the resonant wave-particle interaction effects and the main result is that it leads to a polytropic index $\gamma$ that is complex. The reader is referred to Appendix A for the details about the procedure to include the kinetic effects in the standing problem. The most important conclusion of this analysis is that for the standing situation, the complex polytropic index $\gamma$ is exactly the same as in the single propagating case. This means that a standing wave will show exactly the same attenuation due to LD as the equivalent single propagating wave, at least when there is no initial Alfv\'enic pump in the system. In fact, as we will demonstrate later, the presence of the pump driver does not change the attenuation by Landau damping of the ion-acoustic wave either.

Therefore, as shown by \citet{araneda1998} for the propagating case,  to include the kinetic effects in the standing problem we just need to use the derived dispersion relation of the multi-fluid case and change $\gamma$ of protons by Eq.~(\ref{eq:22}) to obtain a new dispersion relation where $\omega$ is complex and takes into account the attenuation by LD. Hence,
\begin{eqnarray}
    \omega^2=k^2 \left(c^2_{\rm se}+c^2_{\rm sp}\right) = k^2 \left[\gamma_{\rm e} \frac{k_{\rm B}}{m_{\rm p}} T_{\rm e}+2\left(\xi_{s}^2-\frac{1}{Z'(\xi)}\right)\frac{k_{\rm B}}{m_{\rm p}}T_{\rm p}
    \right]. \label{disperrel}
\end{eqnarray}
where $ \xi=\frac{\omega}{k\,w_{\rm 0p}}$,  $w_{\rm 0p}=\sqrt{2\kappa_{\rm B} T_{\rm p}/m_{\rm p}}$ is the thermal velocity of protons and for simplicity we assume that the parallel drift is zero. It is convenient to define $c_{\rm se}=v_{\rm A} \sqrt{\gamma_{\rm e}\,\beta_{\rm e}/2}$ and we have that  $T_{\rm p}/T_{\rm e}=\beta_{\rm p}/\beta_{\rm e}$. From  Eq.~(\ref{disperrel}) is not difficult to write an explicit form for the dispersion relation in terms of the $Z$ function, 
\begin{equation}\label{dispersionk}
{Z'(\xi)}=\frac{2 \,\beta_{\rm p}}{\gamma_{\rm e}\, \beta_{\rm e}}.
\end{equation}
This is exactly the same dispersion relation that is obtained by purely kinematic calculations for parallel propagating ion-acoustic waves based on the Vlasov method \citep[e.g.][]{swanson2003}. It is known that this dispersion relation has an approximate solution when the expansion for large arguments of the $Z$ function is used. Under this approximation the real and imaginary parts of the frequency read
\begin{align}\label{wrdis}
\omega_{\rm R}&=k\, c_{\rm se}\,\sqrt{1+3\frac{\beta_{\rm p}}{\gamma_{\rm e}\, \beta_{\rm e}}},\\
\omega_{\rm I}&=- k\, c_{\rm se}\,\sqrt{\frac{\pi\,\gamma_{\rm e}\, \beta_{\rm e}}{2\, \beta_{\rm p}}}\,\xi^2_{\rm R}\, e^{-\xi^2_{\rm R}},\label{widis}
\end{align}
where $\xi_{\rm R}=\frac{\omega_{\rm R}}{k\, w_{\rm 0p}}$. This dimensionless parameter is found to be independent of the wavenumber and the sound speed using the approximation for $\omega_{\rm R}$ given by Eq.~(\ref{wrdis}). Interestingly, the damping per period defined as $\tau_D/P=\omega_{\rm R}/(2\pi |\omega_{\rm I}|)$ is also independent of $k$ and $c_{\rm se}$ using the previous analytic approximations and the dependence is only on the ratio $\beta_{\rm p}/\beta_{\rm e}$.  It is worth mentioning that LD, apart from producing the attenuation of the oscillation, has a clear effect on the real part of the frequency of oscillation, see Eq.~(\ref{wrdis}), and this change in the period of oscillation might be easier to study in the time-dependent simulations rather than the damping time which is more difficult to estimate  based on the attenuation of the amplitudes with time. The implications of these results in the context of coronal loop oscillations are discussed in Sect.~\ref{discussion}.

\subsection{Hybrid simulations results}\label{secthybridsingle}

We have performed hybrid simulations with the aim of understanding and corroborating the previous analytical results, especially regarding kinetic effects, and as a preparation for the driven case investigated latter. The hybrid code we use is one-dimensional in space but retains all three-components of the electromagnetic field, currents and particle velocities \citep{winskeleroy1985,terasawa1986,horowitz1989,winskeomidi1996}. The electrons are considered as a massless and isothermal fluid, whereas ions are treated kinetically as discrete particles-in-cell (PIC) representing a collisionless plasma with a finite generally anisotropic temperatures, and drifting parallel to the mean magnetic field. The advantage of using this hybrid method is that allows us to resolve the ion
dynamics and to integrate the equations over many ion-cyclotron periods, while neglecting the small temporal and spatial scales of the electron kinetic motions. The particle equations are advanced  in time with a leapfrog method, and the fluid moments are computed with a second-order weighting scheme. The electric and magnetic fields are advanced in time solving the fields via a functional iterative explicit scheme as described by \citet{horowitz1989} and the derivatives in the  gradient, curl and divergence terms are estimated using a fourth-order finite difference scheme. A simple isothermal equation for electrons ($\gamma_{\rm e} = 1$) is used. The code, modified from its original Fourier pseudo-spectral version by A. F.-Vi\~{n}as, has been tested and compared with similar numerical codes in several publications \citep[e.g.][]{aranedaetal2002,xieetal2004,aranedaetal2007,moyaetal2012,manevaetal2013}.

\begin{figure}[!ht]
\includegraphics[width=9.25cm]{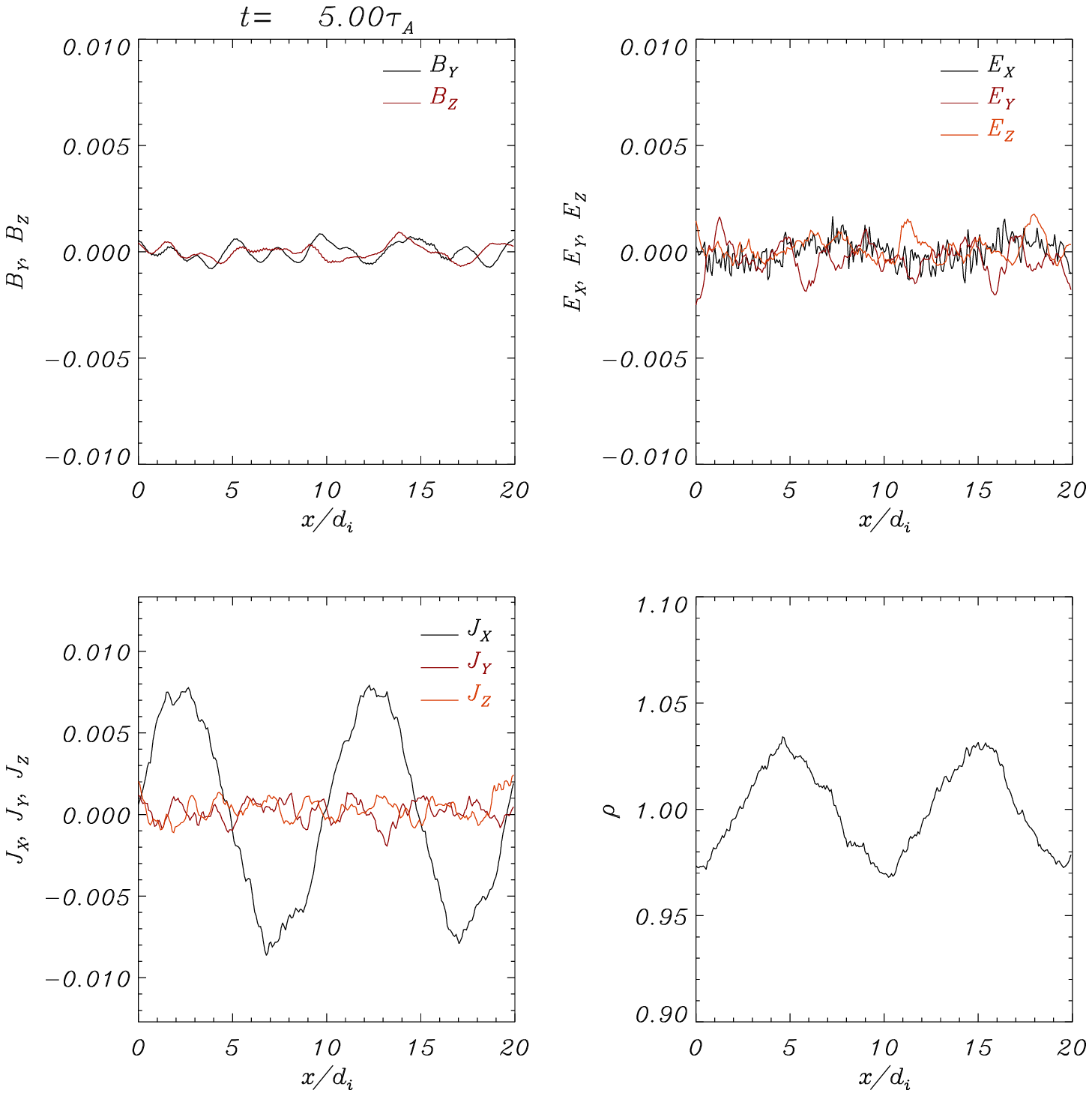}
\includegraphics[width=9.25cm]{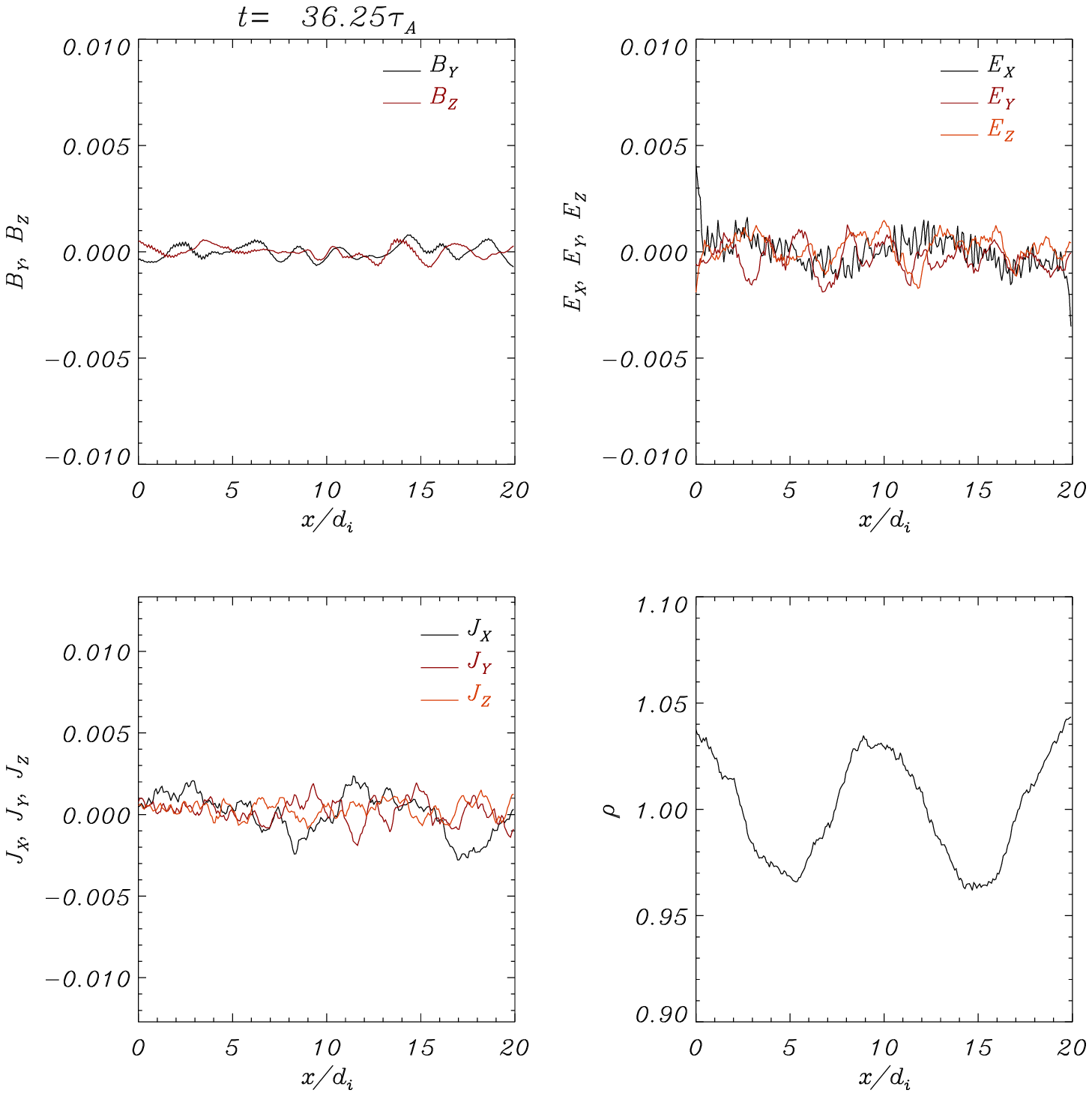}
\caption{\small Top panel, macroscopic variables calculated from the hybrid simulations at $t=5\tau_{\rm A}$.  Wave used $\beta_{\rm e}=6\times 10^{-2}$ and $\beta_{\rm p}=10^{-2}$. Bottom panel, same as the previous panel but at a later stage, $t=36.25\tau_{\rm A}$. A standing ion-acoustic mode is excited in system. In this simulations $L=20d_i$ ($d_i$ is the ion-inertial length) and $k=2\times 2\pi/L$. The magnetic field is normalised as  $B = B_{units}/B_0$, the electric field as $E =c E_{units}/v_{\rm A} B_0$ where $c$ is the speed of light and $v_{\rm A}$ the equilibrium Alfv\'en speed. Density is normalised to the equilibrium density. For visualisation purposes a smoothing of the macroscopic variables has been applied over 10 grid cells. The temporal evolution is available as an online movie.
}\label{figionsnap}
\end{figure}

We use periodic boundary conditions in the system which mimic the effect of line-tying at the edges of the domain. We impose a sinusoidal initial perturbation in the parallel macroscopic velocity (i.e., a perturbation in $\delta u_x$) that excites a harmonic ($k=2\times 2 \pi/L$) of the  ion-acoustic wave  which satisfies the periodic conditions. Hence, we are exciting a single mode in the system which allows a comparison with the previous analytical results. Moreover, the chosen initial mode is precisely the mode that will be excited nonlinearly in the driven case studied in Sect.~\ref{sectdriven}. The macroscopic fluid velocity of the initial perturbation ($\rm \bf u$) is imposed self-consistently with the initial velocities of the individual particles ($\bf {v}$) taking into account the shape of the macro-particles. Since we want to excite a standing mode no density fluctuations are required to be introduced in the system at $t=0$ since density is phase-shifted with respect velocity in time by $90^\circ$ in the standing wave, as we have shown previously. This means that the initial position of the particles does not need to be imposed to obtain a certain initial density profile that excites an ion-acoustic wave. As usual low amplitude random velocities are introduced in the system. We typically use 256 grid cells with 1000 particles per cell. The time step is $\Delta t=0.0025 \, \omega^{-1}_{\rm cp}$. Lengths are normalised to the ion-inertial length, $d_i=v_{\rm A}/\omega_{\rm cp}$ and velocity to the equilibrium Alfv\'en speed, $v_{\rm A}$. We typically use a numerical box length of $L=20 d_i$ although this parameter is not important regarding the damping per period of the ion-acoustic waves. We use that the ratio of the ion plasma frequency to the proton cyclotron frequency is $\omega_{\rm pi}/\omega_{\rm cp}=10^4$.

The time evolution of different macroscopic variables are shown in Fig.~\ref{figionsnap} for a typical run. At the beginning of the simulation (top panel)  the $x-$component of the current shows some coherent spatial structure while the density fluctuation  also displays a large scale associated to the mode number that has initially been excited in the system ($m=4$ in our case). Note that, as expected, $J_x$ (essentially $u_x$) has nodes at the boundaries of the domain while the density has antinodes. The magnetic field does not show a significant amplitude since the noise is low and the same applies to the electric field components except for $\delta E_x$. As time evolves (bottom panel) the oscillatory character of the variables becomes visible and the standing character of the wave is clear. In these simulations we have found a good agreement with the analytical expressions given previously. In particular we have checked that the overall structure of the parallel velocity and density is in agreement with the Eqs.~(\ref{solvpar2vp0}) and (\ref{solvpar2nm}). Although the electric field seems quite noisy in Fig.~\ref{figionsnap}, the longitudinal component behaves also as predicted by Eq.~(\ref{epar0ex}). For a better comparison of the simulations and the multi-fluid solutions for the standing case we have focused on a situation with 
very small Landau damping by choosing a small value of $\beta_{\rm p}$ in comparison with $\beta_{\rm e}$.

\begin{figure}
\includegraphics[width=8.cm]{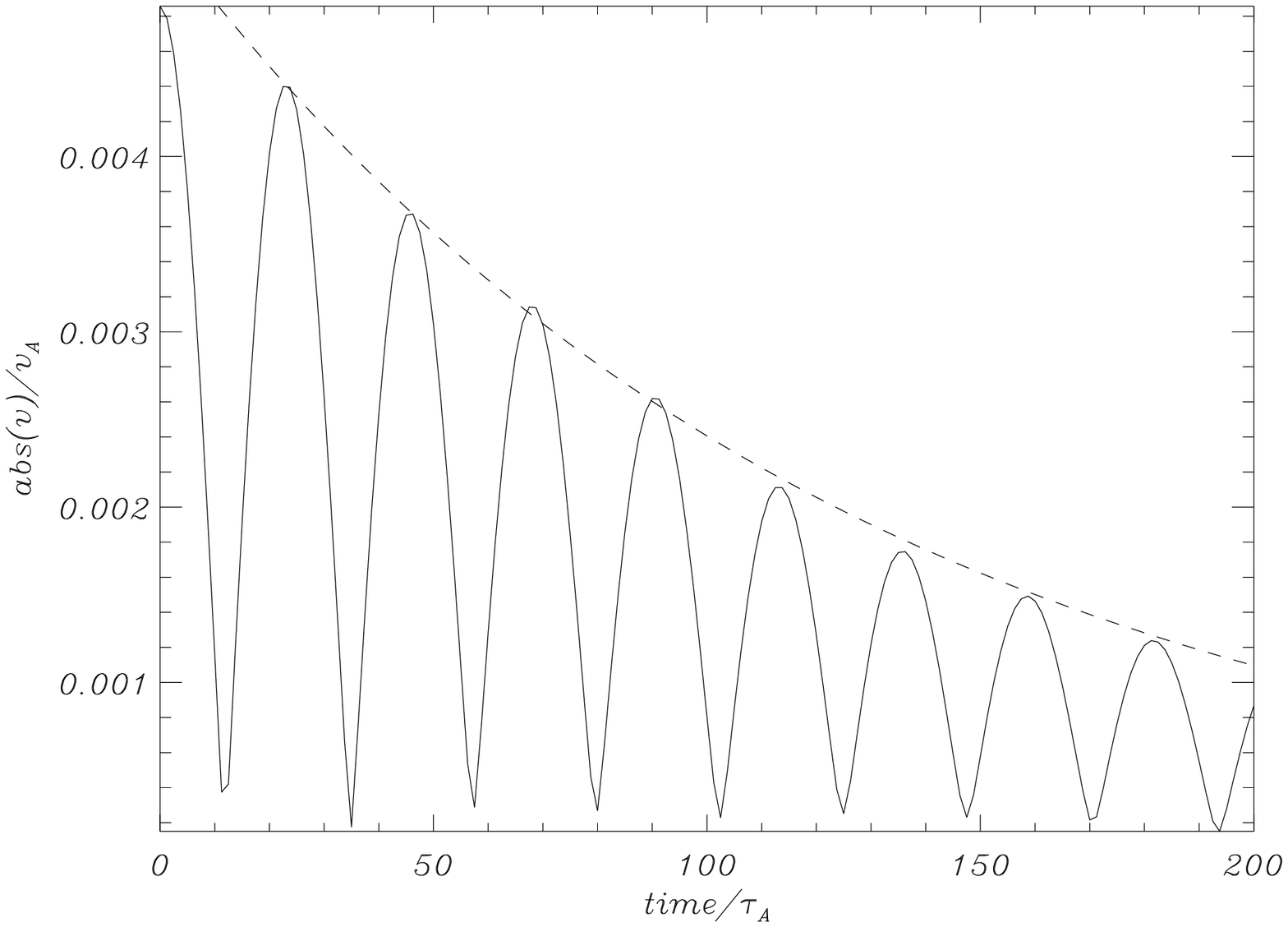}
\includegraphics[width=8.cm]{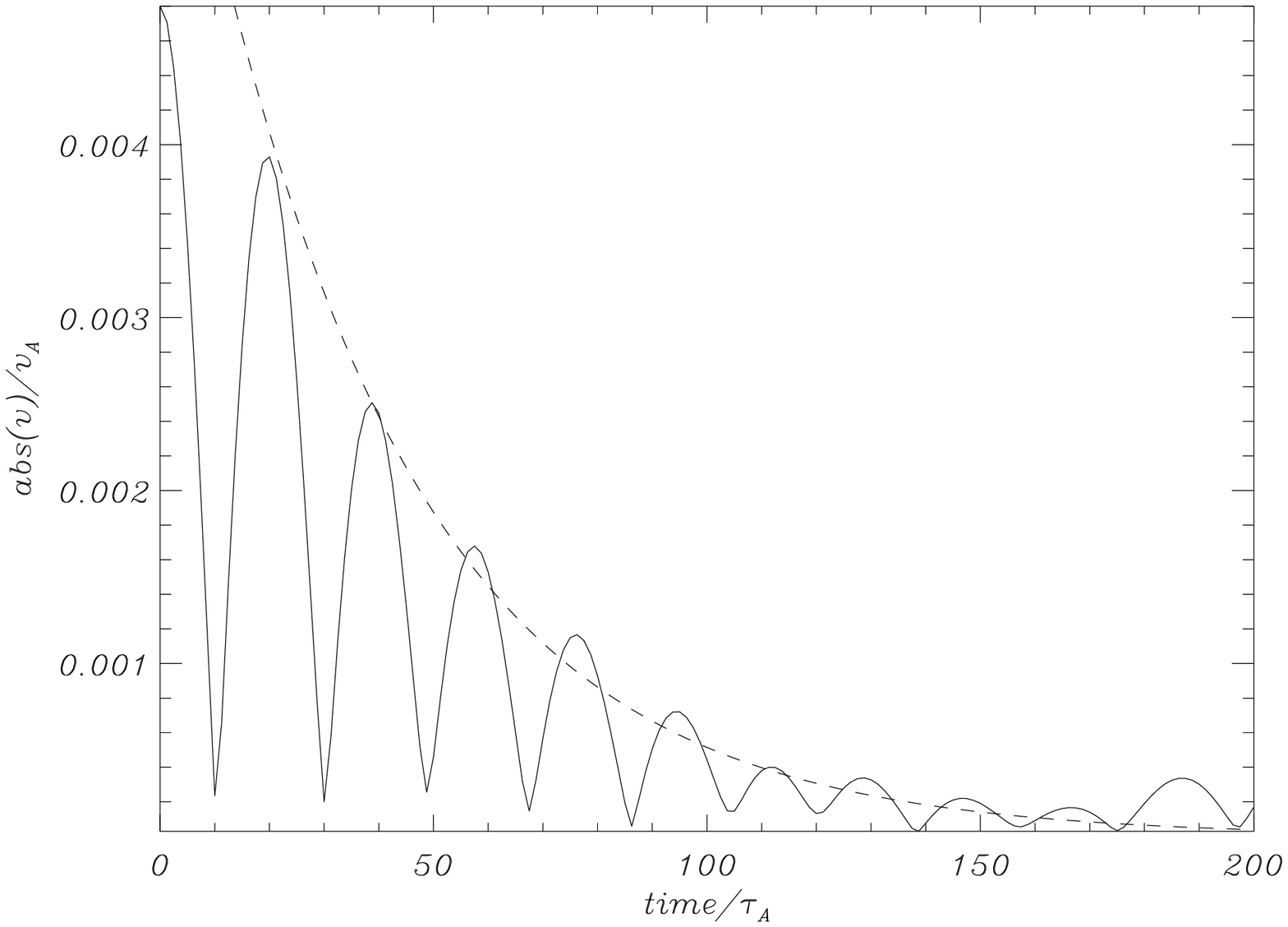}
\caption{\small Top panel, absolute value of the coefficient of the Fourier velocity (macroscopic $u_x$) inferred from the hybrid simulations corresponding to the excited wavelength in the system ($m=4$ or equivalently $m'=2$). The dashed line represents the expected damping from the theoretical calculations based on the solution of the dispersion relation given by Eq.~(\ref{dispersionk}). We have used $\beta_{\rm e}=6\times 10^{-2}$ and $\beta_{\rm p}=10^{-2}$. Bottom panel, same as the previous panel but in this case $\beta_{\rm p}=2\times 10^{-2}$, and therefore the attenuation is stronger.
}\label{figiondamped}
\end{figure}

Next we have considered a situation with strong LD. The results of the simulations indicate that the amplitude of the excited ion-acoustic is attenuated with time. We have inferred the corresponding period and damping time from the simulations. To do so we have performed a Fourier analysis in space and have focused on the dominant mode which is precisely $m=4$ because of the initial excitation. When Fourier analysis is performed the common definition of the wavenumbers is $k=2\pi/L\, m'$, and the mode $m=4$ using Eq.~(\ref{kdis}) corresponds to the mode $m'=2$. An example of the temporal evolution associated to this dominant wavenumber is found in Fig.~\ref{figiondamped} top panel. The damping of the signal is quite clear in this plot, and the expected attenuation due to Landau damping, by solving Eq.~(\ref{dispersionk}), has been also plotted with a dashed envelope for comparison purposes. The results from the dispersion relation are $P=44.6 \tau_{\rm A}$ and $\tau_D=-127.2\tau_{\rm A}$. From the simulations we have performed a periodogram \citep{scargle1982}, providing a better frequency accuracy than the common Fourier power spectrum, to calculate the real part of the frequency. An exponential fit to the envelope of the signal has used to estimate the imaginary part of the frequency. The obtained values from the simulations are $P=45.2\tau_{\rm A}$ and $\tau_D=-124.0\tau_{\rm A}$.  The agreement between the hybrid simulations and the theoretical result is rather good. Another example is shown in the bottom panel of Fig.~\ref{figiondamped}. For the chosen parameters, with twice the plasma beta for protons, the attenuation is faster than in the previous case and the fit to the exponential decay is slightly worse. The results from the dispersion relation are $P=38.1 \tau_{\rm A}$ and $\tau_D=-38.7\tau_{\rm A}$, while the inferred values from the simulations are $P=36.9\tau_{\rm A}$ and $\tau_D=-42.9\tau_{\rm A}$. For large times the amplitude of the signal starts to be dominated by noise.

\begin{figure}
\centering{\includegraphics[width=7.cm]{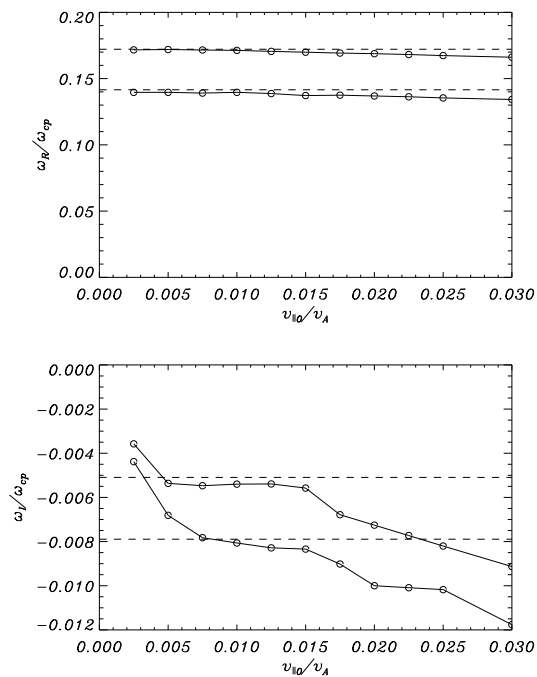}
} 
\caption{\small Real and imaginary parts of $\omega$ for a standing ion-acoustic  wave derived from hybrid simulations (circles) as a function  of the amplitude of the initial excitation. In this example, for the top curve, $\beta_{\rm e}=6\times 10^{-2}$ and $\beta_{\rm p}=10^{-2}$, while for the bottom curve,  $\beta_{\rm e}=10^{-1}$ and $\beta_{\rm p}=1.25\times 10^{-2}$. The attenuation is calculated from the Fourier analysis in space and the power of the mode $m=4$. The horizontal dashed lines corresponds to the expected damping per period of the corresponding non-driven ion-acoustic mode attenuated by LD and calculated from Eq.~(\ref{dispersionk}).
}\label{figiondamperiodion}
\end{figure}

\begin{figure}
\centering{\includegraphics[width=8.cm]{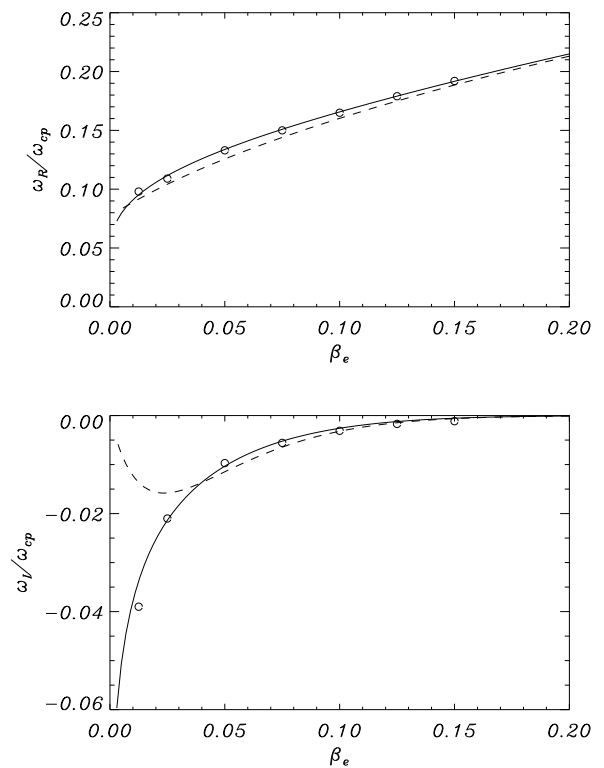}}
\caption{\small Real and imaginary parts of $\omega$ for a standing ion-acoustic  wave derived from hybrid simulations (circles) as a function of the plasma-$\beta$ for electrons. The continuous line corresponds to the numerical solution to the dispersion relation given by Eq.~(\ref{dispersionk}) while the dashed line corresponds to the analytical approximation given by Eqs.~(\ref{wrdis}) and (\ref{widis}). In this plot $\beta_{\rm p}=10^{-1}$, $L=20 d_i$ ($d_i$ is the ion-inertial length) and $k=2\times 2\pi/L$.
}\label{figcompar}
\end{figure}

It is important to mention that the amplitude of the initial wave can have some influence on the attenuation of the signal. Since Landau theory is based on linear results, there might be some effects that are beyond the theoretical predictions. For this reason, we have decided to investigate the dependence of the attenuation with the amplitude of the initial wave using the hybrid simulations. The results are shown in Fig.~\ref{figiondamperiodion} for two particular choice of values of $(\beta_{\rm e},\beta_{\rm p})$. The variation of the real part of the frequency with the amplitude is quite small (top panel). While for the imaginary part, the plot indicates that there is a sort of plateau where the inferred values tend to the theoretical calculations (plotted with horizontal dashed lines). Nevertheless, from Fig.~\ref{figiondamperiodion} we realise that if the initial amplitude is too small or too large the deviations from the expected value can be significant. This is especially relevant for large values of the initial amplitude and the reason behind the faster attenuation is that the excited ion-acoustic wave starts developing shocks. These shocks produce the excitation of higher spatial harmonics and therefore some energy is deposited into these modes, meaning that the energy of the initial $m=4$ mode decreases. Since the numerical code we are using does not contain proper shock-capturing techniques or explicit dissipation this issue is not investigated further and we concentrate on the lineal regime. On the other limit, for very small amplitudes, we have to remember that the system contains random noise that in this case it is able to alter the efficiency of the LD process. We think that the results shown here are not related to the nonlinear effects on the resonance itself. \citet{oneil1965} showed that the damping can be significantly reduced because of the nonlinear energy exchange between a wave and the resonant particles trapped in its potential wells. As the phase mixing of resonant particles becomes more complex, the less energy interchange occurs. In reality it is known that the wave amplitude eventually saturates at a nonzero value which depends on the amplitude of the initial perturbation \citep[e.g.][]{lancelottidoring1998,lancelottidoring2009}.

With the knowledge of the most convenient amplitude (typically around $0.01 v_{\rm A} $) to be in the proper attenuation regime dominated by linear Landau damping we have constructed Fig.~\ref{figcompar} by changing $\beta_{\rm e}$ and keeping the rest of the parameters constant. We have represented again the real and imaginary part of omega together with  the numerical solution of the dispersion relation and the analytical approximations. We find that the results inferred from the simulations agree quite well with the numerical solution for both the real and imaginary parts of $\omega$. The analytical approximation is quite good for the real part of omega while the imaginary part is less accurate \citep[this is common in this kind of approximation, see for example,][]{swanson2003}. The estimation of the damping time is more difficult from the simulations, especially when the attenuation is  fast, but the fact that the real part agrees well with  the theoretical prediction is already a clear  indication that ion-acoustic waves loses their energy due to LD.

%

\begin{figure}
\centering{\includegraphics[width=8.cm]{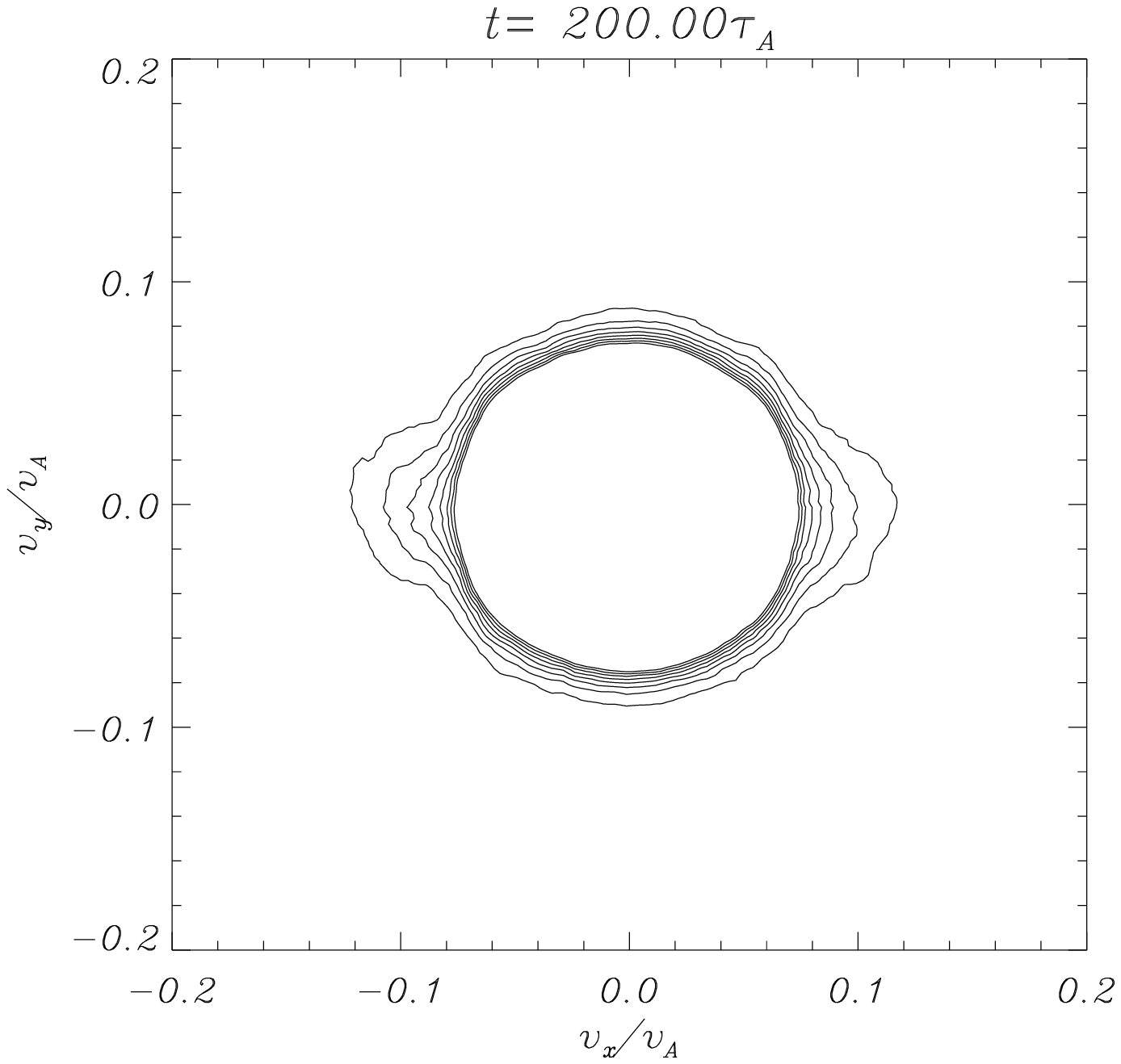}}
\centering{\includegraphics[width=8.cm]{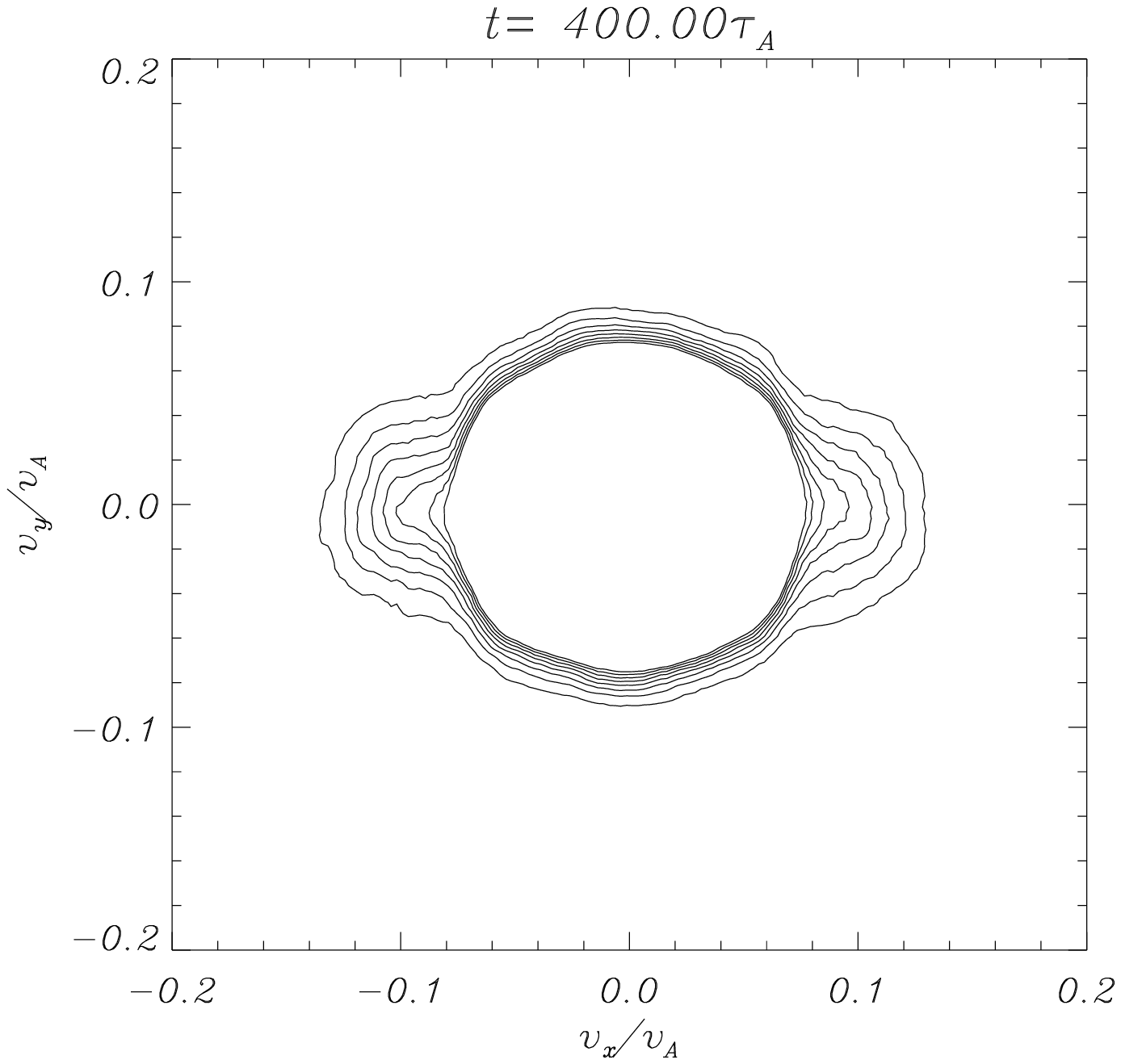}}
\caption{\small Velocity distribution function as a function of $v_x$ and $v_y$ averaged over $x$ and calculated from the 1D hybrid simulations at two different times of the evolution. Top panel corresponds to the initial development of the beams due to LD. Bottom panel shows a later development of the beams. We have used a logarithmic scale to represent the distribution function for a better visualisation of the results. In this plot we have used $\beta_{\rm e}=6\times 10^{-2}$ and $\beta_{\rm p}=10^{-3}$ and corresponds to weak LD.
}\label{figvdf}
\end{figure}


Further evidences of the presence of Landau damping are found by calculating the velocity distribution function (VDF) from the hybrid simulations. In Fig.~\ref{figvdf} we find the VDF as a function of the parallel velocity of the particles, $v_x$, and one of the transverse particle velocity components, $v_y$, integrated over the spatial simulation box along the $x-$direction. Similar results are obtained for the $v_x$--$v_z$ velocity components. The initially circular profile at $t=0$ (not shown in the plot) starts to develop two symmetrical beams  along $v_x$, since the standing wave can be viewed as the superposition of two identical counter-propagating waves. As time advances the beams are enhanced since more energy is deposited around the two symmetric resonance positions due to the wave-particle interaction. For a propagating wave we would only find a single beam instead of two, similar to the results obtained by \citet{aranedaetal2007}.

The conclusion of this section is that standing ion-acoustic waves are eigenmodes of our finite length system and that these waves, similarly to the propagating case, attenuate by LD. In the linear regime the attenuation times and frequencies are exactly the same as those obtained for propagating waves and we can use the analytical approximations known for that situation. The fact that the standing wave is a combination of  two counterpropagating waves does not change the efficiency of the wave-particle interaction process. 

\section{Ion acoustic waves driven by Alfv\'en waves}\label{sectdriven}

The goal here is to investigate the generation of ion-acoustic waves driven nonlinearly due to the presence of a standing Alfv\'enic pump. The previous section, based on the calculation of the non-forced ion-acoustic eigenmodes, turns out to be useful to understand the present driven case. 


Since the Alfv\'enic standing pump can be viewed as  the superposition of two circularly polarised counter-propagating waves, the first question that needs to be addressed is whether parametric instabilities are present in the system. For this reason we think that it is convenient to briefly reexamine the mathematical analysis that leads to the physical description of parametric instability for a circularly polarised propagating Alfv\'en wave. The idea is to use a perturbational approach and derive the equations at different orders in a dimensionless parameter. In this case \citep[see][]{derby1978,goldstein1978} the correct ordering in the perturbation scheme in the ideal MHD equations is (the parameter is  $\epsilon$ and it is assumed to be much smaller than 1)
\begin{align}\label{eq:param}
{\bf u}&={\bf u_{\perp}}+\epsilon\,{\bf u'_{\perp}}+\epsilon\, {\delta u_{x}}\,{\bf {\hat e}_x},\nonumber\\
{\bf B}&=B_0\,{\bf \hat e_x} + {\bf B_{\perp}}+\epsilon\,{\bf B'_{\perp}},\nonumber\\
\rho&=\rho_0+\epsilon\, \delta \rho,\nonumber\\
p&=p_0+\epsilon\,\delta p.
\end{align}



\noindent We start with zero order in $\epsilon$. We obtain from the perpendicular component of the momentum equation that
\begin{eqnarray}\label{vwave0}
\rho_0\frac{\partial {\bf u_{\perp}}}{\partial t}=\frac{B_0}{\mu_0} \frac{\partial {\bf B_{\perp}}}{\partial x},
\end{eqnarray}
while from the induction equation to zero order in $\epsilon$ we find 
\begin{eqnarray}\label{mwave0}
\frac{\partial {\bf B_{\perp }}}{\partial t}=B_0 \frac{\partial {\bf u_{\perp }}}{\partial x}.  
\end{eqnarray}
Combining Eqs.~(\ref{vwave0}) and (\ref{mwave0}) we obtain the standard wave equation that describes Alfv\'en waves.
It is not difficult to see that if the propagating Alfv\'en wave is circularly polarised  (the situation considered here) then $B_{\perp}=\text{const}$, i.e., the amplitude of the wave is independent of time and position (see Appendix C), and this has important consequences as we show in the following. 

The parallel component of the momentum equation leads to zero order, 
\begin{eqnarray}\label{eq:wavecircpol}
0=-\frac{\partial p_0}{\partial x}-\frac{1}{2 \mu_0}\frac{\partial B^2_{\perp}}{\partial x},
\end{eqnarray}
which is a basic equation that is not clearly described in the literature of parametric instabilities,
Since $B_{\perp}=\text{const}$ this means that the second term in the previous equation,  i.e., the ponderomotive term, is zero and therefore $p_0$ must be constant as well and the equation is fulfilled. However, and this is the fundamental point of the present paper, for a standing circularly polarised wave the ponderomotive force of the pump is not zero and a different approach regarding the perturbation scheme needs to be used. But for the moment we still provide the equations to first order in $\epsilon$ for the propagating circularly polarised wave. From the continuity equation we find
\begin{eqnarray}
\frac{\partial \delta \rho}{\partial t}+\rho_0 \frac{\partial {\delta u_x }}{\partial x}=0. 
\end{eqnarray}
The perpendicular component of the momentum equation gives 
\begin{eqnarray}\label{eq:velperpp}
\rho_0\frac{\partial {\bf u'_{\perp}}}{\partial t}+\delta \rho\frac{\partial {\bf u_{\perp}}}{\partial t}+\rho_0 {\delta u_x} \frac{\partial {\bf u_{\perp }}}{\partial x}=\frac{B_0}{\mu_0} \frac{\partial {\bf B'_{\perp}}}{\partial x},
\end{eqnarray}
while for the parallel component
\begin{eqnarray}\label{vpareq0}
\rho_0\frac{\partial {\delta u_x}}{\partial t}=-\frac{\partial \delta p}{\partial x}-\frac{1}{\mu_0} \frac{\partial ({\bf B_{\perp}}\cdot {\bf B'_{\perp}})}{\partial x}. 
\end{eqnarray}
It is important to realise that on the right hand side of this equation we find the coupling between the pump, $\bf B_{\perp}$, and the induced magnetic perturbations, $\bf B'_{\perp}$. From the energy equation we obtain
\begin{eqnarray}
\frac{\partial \delta p}{\partial t}=-\gamma p_0 \frac{\partial {\delta u_{x}}}{\partial x},
\end{eqnarray}
which is usually substituted in Eq.~(\ref{vpareq0}) to eliminate the variable $\delta p$. The magnetic field perturbation to first order in $\epsilon$ is
\begin{eqnarray}\label{eq:bperpp}
\frac{\partial {\bf B'_{\perp}}}{\partial t}=B_0 \frac{\partial {\bf u'_{\perp}}}{\partial x}-\frac{\partial ({\delta u_x} {\bf B_{\perp }})}{\partial x},
\end{eqnarray}
where we see that there is a coupling with the longitudinal motions.
These first order equations in $\epsilon$ are exactly the same as those derived by \citet{derby1978,goldstein1978} although this last author rewrites the equations as a second order derivative in time and space. The same equations are found in  \citet{wonggolstein1986} and \citet{jayantihollweg1993} when dispersion effects, introduced by the finite ion cyclotron, are removed. These last authors, as in \citet{derby1978}, prefer to write the equations in terms of $B_\pm=B_y \pm i B_z$, which is useful when working with circularly polarised waves (the positive sign corresponds to left handed waves and the negative sign to right handed waves). It is well established that the previous  equations lead to parametric instabilities, see the references in this paragraph.

If instead of a circularly polarised propagating wave we consider a standing circularly polarised wave, i.e., the main topic of the present work, we have that $B_{\perp} \neq \text{const}$ (see Appendix C). This means that the perturbation expansion can not be applied essentially because the ponderomotive force due to the pump is not zero now. The previous perturbation scheme fails precisely in Eq.~(\ref{eq:wavecircpol}) since the second term of the right hand side is different from zero. Hence, we have to perform a different analysis to properly include the effect of the ponderomotive force in the equations, missing in the  circularly polarised propagating case.  It turns out that the appropriate expansion when $B_{\perp} \neq \text{const}$ is the following \citep[e.g.][]{rankinetal94, tikhoetal1995, ballesteretal2020},
\begin{align}\label{exp1}
{\bf u}&= \epsilon\,{\bf u_{\perp}}+\epsilon^2\, {\delta u_{x}}\,{\bf {\hat e}_x}+\epsilon^3\,{\bf u'_{\perp}},\nonumber\\
{\bf B}&=B_0\,{\bf {\hat e}_x} + \epsilon\,{\bf B_{\perp}}+\epsilon^3\,{\bf B'_{\perp}},\nonumber\\
\rho&=\rho_0+\epsilon^2\, \delta \rho,\nonumber\\
p&=p_0+\epsilon^2\,\delta p.
\end{align}
This expansion scheme is different from the situation for the circularly propagating case given by Eq.~(\ref{eq:param}). It is therefore logical to find a distinct physical behaviour in the two situations. In particular we realise that using the previous expansion the second order density, pressure and the parallel velocity perturbations do not have the same order as the induced perpendicular velocity and magnetic field fluctuations which are third order.

We use the same procedure and collect terms of the same order in $\epsilon$. From the perpendicular component of the momentum equation we obtain to first order
\begin{eqnarray}\label{vwave0c}
\rho_0\frac{\partial {\bf u_{\perp}}}{\partial t}=\frac{B_0}{\mu_0} \frac{\partial {\bf B_{\perp}}}{\partial x}.
\end{eqnarray}
while the induction equation to first order in $\epsilon$ we find 
\begin{eqnarray}\label{mwave0c}
\frac{\partial {\bf B_{\perp }}}{\partial t}=B_0 \frac{\partial {\bf u_{\perp }}}{\partial x}.  
\end{eqnarray}
Combining Eqs.~(\ref{vwave0c}) and (\ref{mwave0c}) we obtain again the standard wave equation that describes Alfv\'en waves. This is exactly the same result as in the previous propagating case, but now it is found to be to first order in $\epsilon$. Let us provide the equations to second order in $\epsilon$, from the continuity equation we find
\begin{eqnarray}
\frac{\partial \delta \rho}{\partial t}+\rho_0 \frac{\partial {\delta u_x }}{\partial x}=0. 
\end{eqnarray}
From the parallel component of the momentum equation we obtain, to second order in $\epsilon$ that
\begin{eqnarray}\label{vpareq0s}
\rho_0\frac{\partial {\delta u_x}}{\partial t}=-\frac{\partial \delta p}{\partial x}-\frac{1}{2 \mu_0} \frac{\partial { B_{\perp}^2}}{\partial x}. 
\end{eqnarray}
This equation was first derived in \citet{hollweg71} (see his Eq.~(11)) where a definite connection between linearly polarised Alfv\'en waves and density fluctuations was established. 
It is important to realise that {\bf on} the right hand side of this equation we encounter the effect of the ponderomotive force and this equation is different from that obtained in the circularly polarised propagating case to first order, see second term on the right hand side of Eq.~(\ref{vpareq0}). This is the reason that leads to different physical processes, namely a ponderomotive driven system in one case and to a parametrical unstable configuration in the other.

Interestingly, if we continue the analysis, to third order we obtain,
\begin{eqnarray}\label{eq:vfieldstand}
\rho_0\frac{\partial {\bf u'_{\perp}}}{\partial t}+\delta \rho\frac{\partial {\bf u_{\perp}}}{\partial t}+\rho_0 {\delta u_x} \frac{\partial {\bf u_{\perp }}}{\partial x}=\frac{B_0}{\mu_0} \frac{\partial {\bf B'_{\perp}}}{\partial x},
\end{eqnarray}
while for the magnetic field perturbation to third order in $\epsilon$ we have
\begin{eqnarray}\label{eq:bfieldstand}
\frac{\partial {\bf B'_{\perp}}}{\partial t}=B_0 \frac{\partial {\bf u'_{\perp}}}{\partial x}-\frac{\partial ({\delta u_x} {\bf B_{\perp }})}{\partial x}.
\end{eqnarray}
These last two equations are exactly the same as in the circularly polarised propagating case, compare with Eqs.~(\ref{eq:velperpp}) and (\ref{eq:bperpp}). But in the present scheme they are third order in $\epsilon$, while the longitudinal velocity component is second order. This is different to the purely parametric unstable situation where density fluctuations and the induced velocity and magnetic field perturbations are of the same order in $\epsilon$. Nevertheless, Eqs.~(\ref{eq:vfieldstand}) and (\ref{eq:bfieldstand}) can still be viewed as the mathematical representation of a potential parametric process but to third order. Higher order terms can also play a role in regarding parametric instabilities. Thus, the main conclusion here is that for the standing wave problem parametric instabilities are not as primary as in the propagating case, and ponderomotive effects are prevailing. Therefore, we need to focus on the  the effect of the ponderomotive force, and for this reason we extend the previous results, using the perturbational approach given in Eq.~(\ref{exp1})  by using the multi-fluid equations that provide a more general description of a plasma (we assume again that electrons are massless). The electric field, present in the multi-fluid description, is assumed to have a similar expansion as the fluid velocity.

\subsection{First order linearisation}
Collecting terms to first order in $\epsilon$ in the perturbational scheme we find the known equations that describe the circularly Alfv\'enic pump. Since we concentrate on standing waves instead of propagating waves we assume the following dependence (see Appendix C)
\begin{align}\label{valfpol}
{u_\perp} &=i {u_{\perp 0}}\, e^{-i \omega_0 t} \sin{k_0 x},\\
B_\perp &=B_{\perp 0}\, e^{-i \omega_0 t} \cos{k_0 x}.
\end{align}
 Let us briefly describe the features of this wave and how we obtain the dispersion relation and the polarisation relations in the multi-fluid case. The value of the wavenumber $k_0$ is discrete now.

Using Faraday's law the electric field can be written in terms of the magnetic field as \begin{eqnarray}\label{eqEperp}
E_\perp=\frac{\omega_0}{c k_0} B_{\perp 0}\, e^{-i \omega_0 t} \sin{k_0 x}.
\end{eqnarray}
The vectorial product between the velocity and the background is simply $-i B_0 {u_\perp}$. To first order in $\epsilon$ we find 
\begin{eqnarray}\label{eqvperpalf}
-i \omega_0 {u_{\perp 0}}=\frac{q_s}{m_s}\left(\frac{\omega_0}{c k_0} B_{\perp0}+ \frac{B_0}{c} {\delta u_{\perp0}}\right). 
\end{eqnarray}
If we introduce the cyclotron frequency $\omega_{\rm c}=q_s B_0/ m_s c$ (for protons $q_s=e$ and $m_s=m_{\rm p}$ while for electrons $q_s=-e$ and $m_s=m_{\rm e}$) the previous equation is written as
\begin{eqnarray}\label{eqvpol}
\left(\omega_{\rm c}-\omega_0\right) {u_{\perp 0}}=-\omega_{\rm c} \frac{\omega_0}{k_0}  \frac{B_{\perp 0}}{B_0}. 
\end{eqnarray}
It is important to note that this equation is valid for both protons and electrons and they have different velocity amplitudes in the perpendicular direction during the periodic oscillation. In the case of electrons (assuming that $|\omega_{\rm ce}| \gg \omega_0$) the relation reduces to  
\begin{eqnarray}\label{eqveperp}
{u_{\perp 0 \rm e}}=-\frac{\omega_0}{k_0}  \frac{B_{\perp0}}{B_0}. 
\end{eqnarray}
To derive a dispersion relation we have to use Ampere's law which reads
\begin{eqnarray}\label{ampere0}
j_\perp=-i\frac{c}{4\pi}k_0 B_{\perp 0} \, e^{-i \omega_0 t} \sin{k_0 x},
\end{eqnarray}
where we need the total perturbed current due to protons and electrons which is
\begin{eqnarray}\label{jperp}
j_{\perp}=e n_{\rm 0p} {u_{\perp \rm  p}}-e n_{0\rm e} {u_{\perp \rm  e}}. 
\end{eqnarray}
Due to charge neutrality the density of protons and electrons must be the same, i.e., $n_{\rm 0p}=n_{0\rm e}$. Eliminating the velocities if favour of the magnetic field perturbation  we have
\begin{eqnarray}\label{jperp1}
j_{\perp}=-e n_0 \frac{B_\perp}{k_0 B_0}\frac{\omega_0^2}{\omega_{\rm cp}-\omega_0}.
\end{eqnarray}
Substituting this expression in Eq.~(\ref{ampere0}) we obtain the dispersion relation for standing waves which is exactly the same as that for propagating circularly polarised waves,
\begin{eqnarray}\label{disperalf0}
1-\frac{\omega_0}{\omega_{\rm cp}}=\frac{\omega_0^2}{\omega_{\rm cp}^2}
\frac{\omega_{\rm cp}^2}{k_0^2 v_{\rm Ap}^2}.
\end{eqnarray}
where $v_{\rm Ap}^2=B_0^2/(4\pi n_{\rm 0p} m_{\rm p})$. Defining $X_0=\omega_0/\omega_{\rm cp}$, $Y_0=k_0 v_{\rm Ap}/\omega_{\rm cp}$ the previous equations reduce to the standard form \citep[e.g.][]{gomberoffelgueta1991,cramer2001}
\begin{eqnarray}\label{disperalfm}
Y_0^2=\frac{X_0^2}{1-X_0}.
\end{eqnarray}
For $\omega_0>0$ this equation describes left handed polarised Alfv\'en waves propagating in the positive direction (if $k_0>0$) or in the negative  direction (if $k_0<0$) along the magnetic field, both with the same frequency. 

\begin{figure}
\centering{\includegraphics[width=8.5cm]{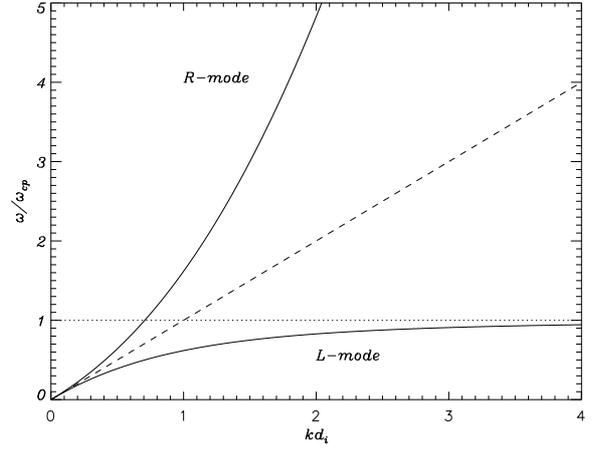}
} 
\caption{\small Dispersion relation for circularly polarised Alfv\'en waves calculated using Eqs.~(\ref{disperalfm}) and (\ref{disperalfml}). The right-handed mode or $R-$mode has a frequency that is unbounded with the wavenumber $k$. The left-handed mode or $L-$mode in the massless electron model considered here has a frequency that tends asymptotically for large $k$ to the ion-cyclotron frequency, $\omega_{\rm cp}$. The dashed line represents the results of the single fluid model and in this case the Alfv\'en wave has a frequency given by $\omega=k v_{\rm A}$. In the limit of very small numbers the wave frequencies of the single and multi-fluid descriptions coincide. 
}\label{figdisperalf}
\end{figure}

For right handed waves we need to change $\omega_0$ by $-\omega_0$ and the dispersion relation is
\begin{eqnarray}\label{disperalfml}
Y_0^2=\frac{X_0^2}{1+X_0},
\end{eqnarray}
and again the same frequency is obtained for forward or backward waves since the dispersion relation is quadratic in $k_0$. Hence, we have obtained a dispersion relation to first order for standing circularly polarised Alfv\'en waves which is exactly the same as for circularly polarised Alfv\'en propagating parallel to the magnetic field. The properties of the dispersion relation for $L$ and $R-$modes are shown in Fig.~\ref{figdisperalf}.

To finish this section it is interesting to realise that the polarisation relation given by Eq.~(\ref{eqvpol}) can be further simplified using the dispersion relation given by Eq.~(\ref{disperalf0}). We finally get that
\begin{eqnarray}\label{eqvperpbperpprot0}
{u_{\perp \rm p}}=-\frac{v_{\rm A}^2}{\omega_0/k_0} \frac{B_\perp}{B_0}. 
\end{eqnarray}
for protons, while for electrons
\begin{eqnarray}\label{eqvperpbperpe0}
{u_{\perp \rm e}}=-\frac{\omega_0}{k_0} \frac{B_\perp}{B_0}. 
\end{eqnarray}
Equations~(\ref{eqvperpbperpprot0}) and (\ref{eqvperpbperpe0}) are valid for both right and left-handed circularly polarised modes but $\omega_0$ changes for $R$ and $L-$modes. 

Note that protons and electrons have the same velocity in MHD, i.e., when $\omega_0=\pm k_0 v_{\rm A}$, and in this case
\begin{eqnarray}\label{eqvperpbperpmhd1}
{u_\perp}=-v_{\rm A} \frac{B_\perp}{B_0}. 
\end{eqnarray}
for $k_0>0$ and
\begin{eqnarray}\label{eqvperpbperpmhd2}
{u_\perp}=v_{\rm A} \frac{B_\perp}{B_0}. 
\end{eqnarray}
for $k_0<0$.

\subsection{Second order linearisation}\label{secsecodord}

To second order in $\epsilon$ we expect to obtain the dynamics of the wave due to ponderomotive effects. Using the standing Alfv\'enic pump to first order, described in the previous section, we have that for the parallel velocity component (same for protons and electrons) and to second order in $\epsilon$
\begin{eqnarray}\label{vpareq}
m_s\frac{\partial {\delta u_x}}{\partial t}=q_s \left(\delta E_x+\frac{1}{c}({\bf u_{s}} \times {\bf B})_{2x} \right)-\frac{1}{n_{0s}}\frac{\partial \delta p_s}{\partial x},
\end{eqnarray}
where to second order,
\begin{eqnarray}
({\bf u_{s}} \times {\bf B})_{2x}=\frac{1}{2i} \left(-u_{\perp s} B^*_{\perp}+u^*_{\perp s} B_{\perp}\right),
\end{eqnarray}
and protons and electrons have different perpendicular velocity amplitudes in the circularly polarised pump (see previous section for the relationship between $u_{\perp s}$ and  $B_{\perp}$).

As usual we eliminate the longitudinal component of the parallel electric field by assuming massless electrons, therefore from Eq.~(\ref{vpareq})  applied to electrons and neglecting the term on the left hand side, we obtain 
\begin{eqnarray}\label{epar}
\delta E_x=-\frac{1}{c}({\bf u_{e}} \times {\bf B})_{2x}-\frac{1}{e n_{0}}\frac{\partial \delta p_e}{\partial x}. 
\end{eqnarray}
This expression is then introduced in Eq.~(\ref{vpareq}) but applied to protons.

Due to quasi-neutrality $n_{\rm 0p}=n_{0\rm e}=n_0$ meaning also that $\delta n_{\rm p}=\delta n_{\rm e}$. Using the standard adiabatic law for the perturbations we have
\begin{equation}\label{eqdensp1}
\frac{\delta p_{s}}{p_{0s}}=\gamma \frac{\delta n_{s}}{n_{0s}}.
\end{equation}
We use this equation to eliminate the pressure perturbation  and to work with the density perturbation.

The resulting system of equations using mass continuity reduces to
\begin{align}\label{vpareqs}
\frac{\partial {\delta u_x}}{\partial t}&=\frac{e}{m_{\rm p}}\frac{1}{c} \left({(\bf u_{\rm p}} \times {\bf B})_{2x}-({\bf u_{e}} \times {\bf B})_{2x} \right)-\left(c^2_{\rm se}+c^2_{\rm sp}\right)\frac{1}{n_{0}}\frac{\partial \delta n}{\partial x},\\
\frac{\partial \delta n}{\partial t}&=-n_0 \frac{\partial {\delta u_x}}{\partial x}.\label{vpareqs1}
\end{align}
Where we have used the definition of the sound speed, $C_{\rm s}$.

We rewrite Eq.~(\ref{vpareqs}) as a single equation for $\delta u_x$,
\begin{align}\label{vpareq2}
\frac{\partial^2 \delta u_x}{\partial t^2}&-\left(c^2_{\rm se}+c^2_{\rm sp}\right)\frac{\partial^2 \delta u_x}{\partial x^2} = \frac{e}{m_{\rm p}}\frac{1}{c}\frac{\partial}{\partial t} \left({(\bf u_{\rm p}} \times {\bf B})_{2x}-({\bf u_{e}} \times {\bf B})_{2x} \right).
\end{align}
We see that we obtain a wave equation with a forcing term on the right hand side due to the Alfv\'enic pump. 

For the density we find the following equation
\begin{align}\label{vpareq2dens}
\frac{\partial^2 \delta n}{\partial t^2}&-\left(c^2_{\rm se}+c^2_{\rm sp}\right)\frac{\partial^2 \delta n}{\partial x^2} =-\frac{n_0}{m_{\rm p}}\frac{e}{c}\frac{\partial}{\partial x} \left({(\bf u_{\rm p}} \times {\bf B})_{2x}-({\bf u_{e}} \times {\bf B})_{2x} \right).
\end{align}
The term on the right hand size involves now the spatial derivative of the forcing term instead of the time derivative that appears in the equation for the density fluctuation. Let us evaluate this forcing term. Using Eq.~(\ref{valfpol}) for the standing pump it is not difficult to find that
\begin{align}\label{forcing}
\frac{e}{m_{\rm p}}\frac{1}{c}& \left({(\bf u_{\rm p}} \times {\bf B})_{2x}-({\bf u_{e}} \times {\bf B})_{2x} \right)
=\nonumber \\ &\frac{e}{m_{\rm p}}\frac{1}{c} \left(u_{\perp \rm 0e}-u_{\perp \rm 0p}\right)B_{\perp 0}\, \sin{k_0 x}  \cos{k_0 x}= \nonumber \\
& \frac{\omega_{\rm cp}}{k_0} \frac{\omega^2_0}{\omega_{\rm cp}-\omega_0} \frac{B^2_{\perp 0}}{B^2_0}\, \frac{1}{2}\sin{2 k_0 x}=A\,f(x),
\end{align}
where we have written the velocity amplitudes for electrons and protons due to the pump wave in terms of the magnetic field perturbation. We rewrite the previous expression as a constant, $A$, times a function, $f(x)$. The forcing term is proportional to the square of the amplitude of the pump, as expected for a second order expansion. However note that there is no temporal dependence. This means that the right hand-side of Eq.~(\ref{vpareq2})  is zero and the parallel motions seem to be uncoupled from the pump. Nevertheless, the forcing term is not zero for the density perturbation, and for this reason we concentrate on this magnitude. We aim at solving Eq.~(\ref{vpareq2dens}) for the forcing term given by Eq.~(\ref{forcing}) together with the following initial and boundary conditions
\begin{align}\label{vcond}
\delta n(x,t=0)&=\frac{\partial \delta n}{\partial t}(x,t=0)=0,\\
\frac{\partial \delta n}{\partial x}(0,t)&=\frac{\partial \delta n}{\partial x}(L,t)=0.
\end{align}
It is known that the formal solution of this problem, is the following 
\begin{align}\label{solvpar}
\delta n(x,t)=\sum^\infty_{m=1} \left[\frac{L}{C_{\rm s}\,m\, \pi}
\int^t_0 f_m\, \sin\left( \frac{C_{\rm s}\,m\, \pi} {L}(t-\tau)\right )d \tau \right] \cos \left( {\frac{m \pi} {L} x}\right) ,\nonumber \\
\end{align}
where
\begin{eqnarray}\label{solvparfn}
f_m=\frac{2}{L}\int^L_0 g(x) \cos\left({\frac{m \pi} {L} x}\right)  dx, \mbox{ } m=1,2,\ldots
\end{eqnarray}
and now we have that
\begin{eqnarray}\label{solvpargx}
g(x)=-n_0\,A\, \frac{\partial f(x)}{\partial x},
\end{eqnarray}
and we have used that $C_{\rm s}=\sqrt{c^2_{\rm se}+c^2_{\rm sp}}$. Substituting $f(x)$ into Eq.~(\ref{solvpargx}) and evaluating the integral in Eq.~(\ref{solvparfn}) we find, due to symmetry reasons, that all the values are zero except $m=4$ (or $m'=2$) obtaining that 
\begin{eqnarray}
f_4=-n_0\,\omega_{\rm cp}\, \frac{\omega^2_0}{\omega_{\rm cp}-\omega_0} \frac{B^2_{\perp 0}}{B^2_0}.
\end{eqnarray}
Using this expression and performing the corresponding integral with respect to $\tau$ in Eq.~(\ref{solvpar}) we eventually obtain the following rather simple dependence for the perturbed density
\begin{align}\label{solvpar20}
\delta n(x,t)=n_0\frac{\omega_{\rm cp}}{(C_{\rm s}\, 2 k_0)^2} \frac{\omega^2_0}{\omega_{\rm cp}-\omega_0} \frac{B^2_{\perp 0}}{B^2_0}\bigg(\cos\left(2 k_0 C_{\rm s} \,t\right)-1\bigg) \cos(2 k_0\, x).\nonumber\\
\end{align}
This equation indicates that the characteristic wavenumber of the induced  perturbations is $2 k_0$ and the corresponding frequency of the ion-acoustic driven wave is $2 k_0 C_{\rm s} $ and it is independent of the frequency of the driver, $\omega_0$. We can simplify further this expression using the dispersion relation given by Eq.~(\ref{disperalf0}) in the multiplicative factor, obtaining
\begin{eqnarray}\label{solvpar2}
\delta n(x,t)=n_0\frac{v^2_{\rm Ap}}{4 C^2_{\rm s}} \frac{B^2_{\perp 0}}{B^2_0}\bigg(\cos\left(2 k_0 C_{\rm s}  \,t\right)-1\bigg) \cos(2 k_0\, x).
\end{eqnarray}

\noindent Solving the wave equation for the velocity, using Eq.~(\ref{vpareqs1}) and  Eq.~(\ref{solvpar2}) it is straight-forward to find
\begin{eqnarray}\label{solvpar2vp}
{\delta u_x}(x,t)=\frac{v^2_{\rm Ap}}{4 C_{\rm s}} \frac{B^2_{\perp 0}}{B^2_0} \sin\left(2 k_0 C_{\rm s} \,t\right)\sin(2 k_0\, x).
\end{eqnarray}
The previous expressions indicate a quadratic dependence of the density and velocity fluctuation with the perturbed perpendicular magnetic field. But recall that for a given perpendicular magnetic fluctuation associated to the Alfv\'enic wave the velocity amplitude for left and right-handed modes is different according to, for example, Eq.~(\ref{eqvperpbperpprot0}).

It is worth mentioning that the previous result is for circularly polarised waves. For linearly polarised Alfv\'en waves an analytical solution is also available and it is different from the one obtained here, as expected. It has been derived using the D’Alembert’s method \citep[see][]{rankinetal94, tikhoetal1995, ballesteretal2020}. In this last case the parallel velocity is the sum of two sinusoidal terms with angular frequencies $2 k_0 c_{s}$ and $2\, \omega_0$. Considering the low $\beta$
regime, $c_{s}\, k_0 \ll  \omega_0$, then the dominant angular frequency of the solution is also $2 k_0 c_{s}$.

\begin{figure}
\centering{\includegraphics[width=9.cm]{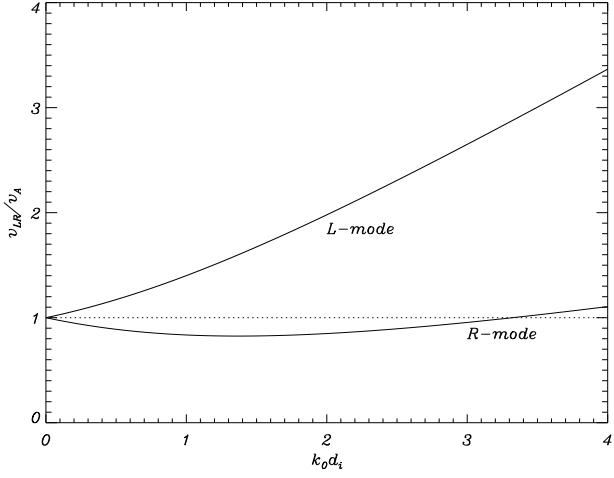}
} 
\caption{\small Modulus of the velocity term, $v_{LR}$, for left and right-handed waves. For the $R-$mode the amplitude of $v_{LR}$ is typically bounded with values around one, while for the $L-$mode it grows with the wavenumber. Again in the limit of very small numbers the wave frequencies of the single and multi-fluid descriptions coincide. In this plot we have used $\beta_{\rm e}=6\times 10^{-2}$ and $\beta_{\rm p}=10^{-2}$.
}\label{figvld}
\end{figure}

It is useful to calculate an expression for the perturbed longitudinal electric field using Eq.~(\ref{epar}) once we know the solution for the density perturbation, $\delta n(x,t)$. We find that
\begin{align}\label{eparf}
\delta E_x(x,t)=&-\frac{1}{c}\frac{B^2_{\perp 0}}{B_0}\frac{1}{2}\left(\frac{\omega_0}{k_0} +
\frac{c^2_{\rm se}}{C^2_{\rm s}}\frac{k_0}{\omega_{\rm cp}} v^2_{\rm Ap}
\right) \sin (2 k_0 x)
\nonumber \\ 
&+\frac{1}{c}\frac{B^2_{\perp 0}}{B_0}\frac{1}{2} \frac{c^2_{\rm se}}{C^2_{\rm s}}
\frac{k_0}{\omega_{\rm cp}} v^2_{\rm Ap}\cos\left(2 k_0 C_{\rm s} \,t\right)\sin(2 k_0\, x).
\end{align}
Hence, the parallel electric field has a time independent contribution (first term) plus another term that accounts for the ion-acoustic oscillation with time. At $t=0$ the previous expression leads to
\begin{eqnarray}\label{eparft0}
\delta E_x(x,t=0)=-\frac{1}{c}\frac{B^2_{\perp 0}}{B_0}\frac{1}{2} \frac{\omega_0}{k_0}\sin (2 k_0 x).
\end{eqnarray}
which is independent of the sound speeds and it only depends on the parameters of the Alfv\'en wave. At later stages of the evolution the generation of density fluctuations modifies the initial parallel electric field according to Eq.~(\ref{eparf}). But let us assume that there is a mechanism that attenuates the ion acoustic oscillation, for example LD. If this is the case this type of oscillations will decay with time and its amplitude will tend to zero for $t\rightarrow \infty$ but the Alfv\'enic wave will continue oscillating. We can use the previous expressions to find the ``static state'' of the system. It is clear that if the ion-acoustic oscillation is attenuated with time then according to Eq.~(\ref{solvpar2vp}) it will eventually lead to $\delta u_x(x,t)=0$. But density is no longer uniform in the asymptotic state and  using Eq.~(\ref{solvpar2}) it is found that
\begin{eqnarray}\label{solvpar2stat}
\delta n(x, t\rightarrow \infty)=-n_0\frac{v^2_{\rm Ap}}{4 C^2_{\rm s}}\frac{B^2_{\perp 0}}{B^2_0} \cos(2 k_0\, x).
\end{eqnarray}
The parallel electric field in this limit accounts only for the first term in Eq.~(\ref{eparf})
\begin{align}\label{eparfstat}
\delta E_x(x,t\rightarrow \infty)&=-\frac{1}{c}\frac{B^2_{\perp 0}}{B_0}\frac{1}{2}\left(\frac{\omega_0}{ k_0} +
\frac{c^2_{\rm se}}{C^2_{\rm s}}\frac{k_0}{\omega_{\rm cp}} v^2_{\rm Ap}
\right) \sin \left(2 k_0 x\right)\nonumber \\ 
&=-\frac{1}{c}\frac{B^2_{\perp 0}}{B_0}\frac{1}{2} {\delta v_{LR}}  \sin \left(2 k_0 x\right).
\end{align}
This means that there is a net parallel electric field in the system created by the Alfv\'enic pump that is able to sustain the created density cavities. These expressions provide information on how the background has changed once the induced ion-acoustic oscillations have disappeared due to some damping mechanism, for example LD. We have defined the velocity $\delta v_{LR}$ as the term inside the parenthesis and this velocity is different for left and right-handed modes. In Fig.~\ref{figvld} we see the dependence of $\delta v_{LR}$ with the wavenumber for certain choice of values of  plasma-$\beta$ for electrons and protons. We conclude that the left-handed mode shows larger parallel electric field fluctuations (because $v_{LR}$ is larger) than right-handed modes. The reason is that if $\omega_0$ is positive for left-handed waves, but negative for right-handed waves then the addition of the two terms in the definition of $v_{LR}$ (the second term is always positive) leading always to a larger value for the $L-$mode.

Hence, we have solved analytically the problem of the nonlinear excitation of the ion-acoustic due to standing circularly polarised Alfv\'enic pump in the absence of LD. Collisionless ion damping is addressed in a similar manner as we have done in Sect.~\ref{kinetion}. The details of the formal derivation are found in Appendix D. The main result is that we can reproduce exactly the same procedure to introduce a kinetic effect in the multi-fluid systems via the polytropic relation between the pressure to second order and the density to second order. In other words, it is justified to substitute again the sound speed by its complex version that includes the effects of Landau damping.

Let us apply the results of the second order analysis to the standing case. Since we have seen that the pump simply excites the first harmonic with wavenumber $2 k_0$  of all the possible ion-acoustic eigenmodes, intuitively we expect that the frequency and damping times are precisely those found in Sect.~\ref{kinetion} for $2 k_0$. But let us try to justify this result by assuming that $\omega$ is complex and substituting the polytropic $\gamma_{\rm p}$ for the corresponding complex expression involving the $Z$ function. According to the analytical results in the absence of LD, we can assume that the velocity is of the form
\begin{eqnarray}\label{vparsimpl}
\delta u_x=A i\, e^{-i\omega t} \sin (2k_0 x),
\end{eqnarray}
where $\omega$ can be now a complex number and $A$ is a constant. From the continuity equation, given by Eq.~(\ref{vpareqs1}) and using Eq.~(\ref{vparsimpl}) we integrate with respect to time obtaining that 
\begin{eqnarray}\label{ndenssimpl}
\delta n=n_0 A i \, 2 k_0 \cos(2 k_0 x) \frac{1}{i \omega}\left(e^{-i\omega t}-1\right).
\end{eqnarray}
This expression is very similar to Eq.~(\ref{solvpar2}) but we have not  determined the value of $\omega$ yet. To derive a dispersion relation we have to use the wave equation for the density given by 
Eq.~(\ref{vpareq2dens}) that includes the forcing term. But if we want to include the  LD effects we have already shown that we have to use the complex version of $C_{\rm s}$ (complex $\gamma_{\rm p}$), defined in Eq.~(\ref{disperrel}). From the density wave equation we find, after cancelling factors that contain the spatial dependence, the following simple equation
\begin{align}\label{wavedensterms}
\frac{1}{i\omega} 2 k_0\, A i \,(-i \omega)^2 e^{-i \omega t}&+C^2_{\rm s} \frac{1}{i\omega} (2 k_0)^3\, A i \left(e^{-i \omega t}-1\right)=-k_0^2 v_{\rm Ap}^2 \frac{B^2_{\perp 0}}{B^2_0}.
\end{align}
This equation has two oscillatory terms proportional to $e^{-i\omega t}$ and two non-oscillatory terms. To make the sum of the two oscillatory terms zero, because the equation must be valid for any time, we find the following condition must be satisfied, 
\begin{eqnarray}\label{disperLD}
\omega=2 k_0\, C_{\rm s},
\end{eqnarray}
where $C_{\rm s}$ depends on $\omega$ now and  contains the $Z$ function. This equation is precisely the dispersion relation for ion-acoustic waves that undergo LD. The corresponding wavenumber is twice that of the Alfv\'enic driver.

In addition, the non-oscillatory term with time on the left hand-side of Eq.~(\ref{wavedensterms}) must be equal to the term on the right hand-side, meaning that 
\begin{eqnarray}\label{constA}
A=\frac{v^2_{\rm Ap}}{4\, C_{\rm s}}\frac{B^2_{\perp 0}}{B^2_0},
\end{eqnarray}
where we have used Eq.~(\ref{disperLD}) to eliminate $\omega$ from the equation. Since we are working with complex numbers we have to take the real part of Eqs.~(\ref{vparsimpl}) and (\ref{ndenssimpl}). If we assume that $C_{\rm s}$ is real (and therefore independent of $\omega$) we recover the results of Sect.~\ref{secsecodord}, compare with  the amplitude in Eq.~(\ref{solvpar2vp}).

The main conclusion is that we obtain exactly the same dispersion relation as in 
Sect.~\ref{kinetion} for non-driven ion-acoustic waves. In the driven case we simply have to change $k$ by $2 k_0$ and solve the dispersion relation or use the analytical expressions for the real and imaginary parts of the frequency already derived. The Alfv\'enic pump, at second order in $\epsilon$, does not modify the attenuation by Landau damping of the generated ion-acoustic waves. The reason is that the forcing term on the right-hand side of Eq.~(\ref{wavedensterms}) does not contain any dependence with the sound speed, $C_{\rm s}$, that accounts for LD effects.

\begin{figure}[!ht]
\centering{\includegraphics[width=9.25cm]{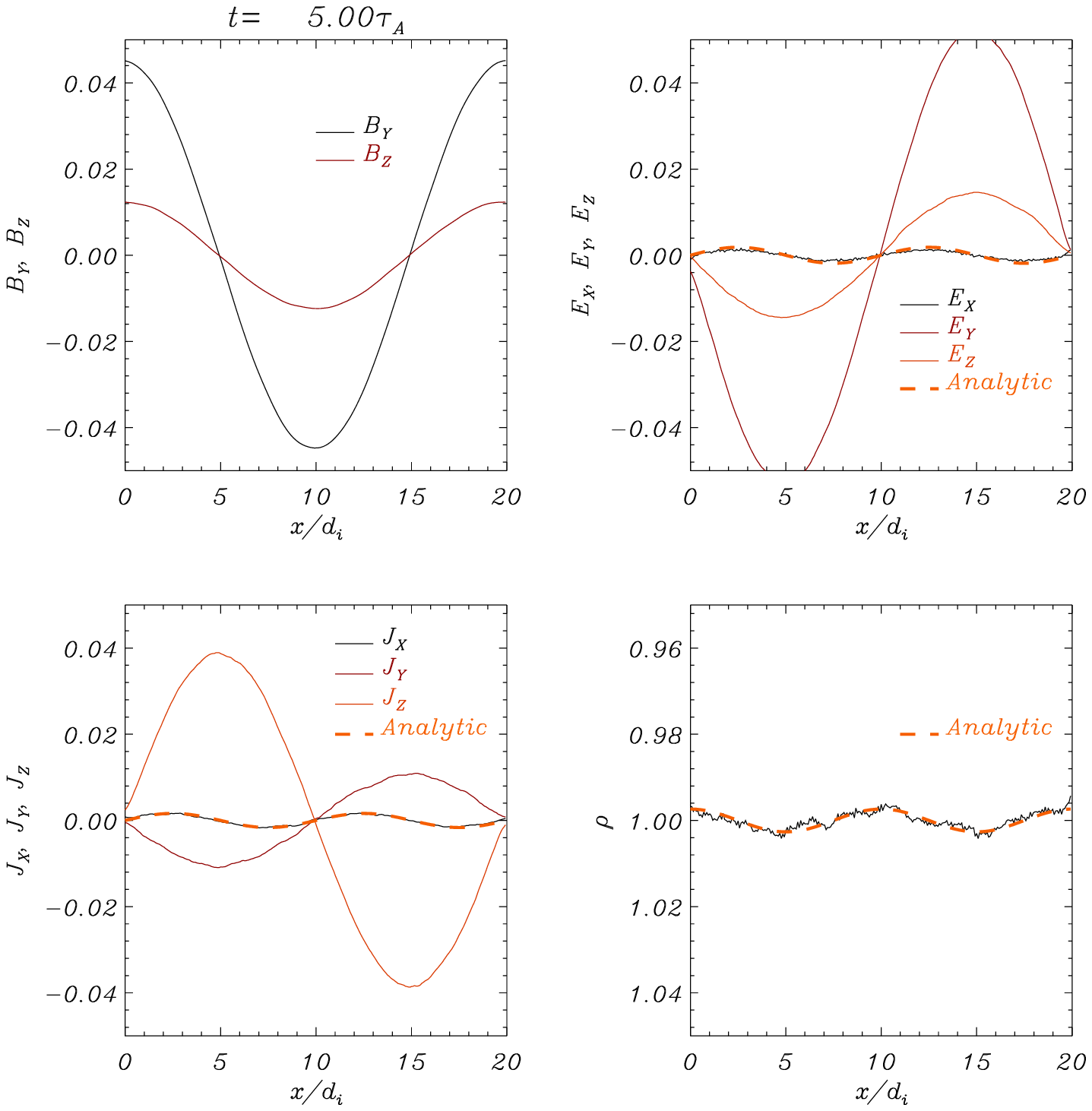}
\includegraphics[width=9.25cm]{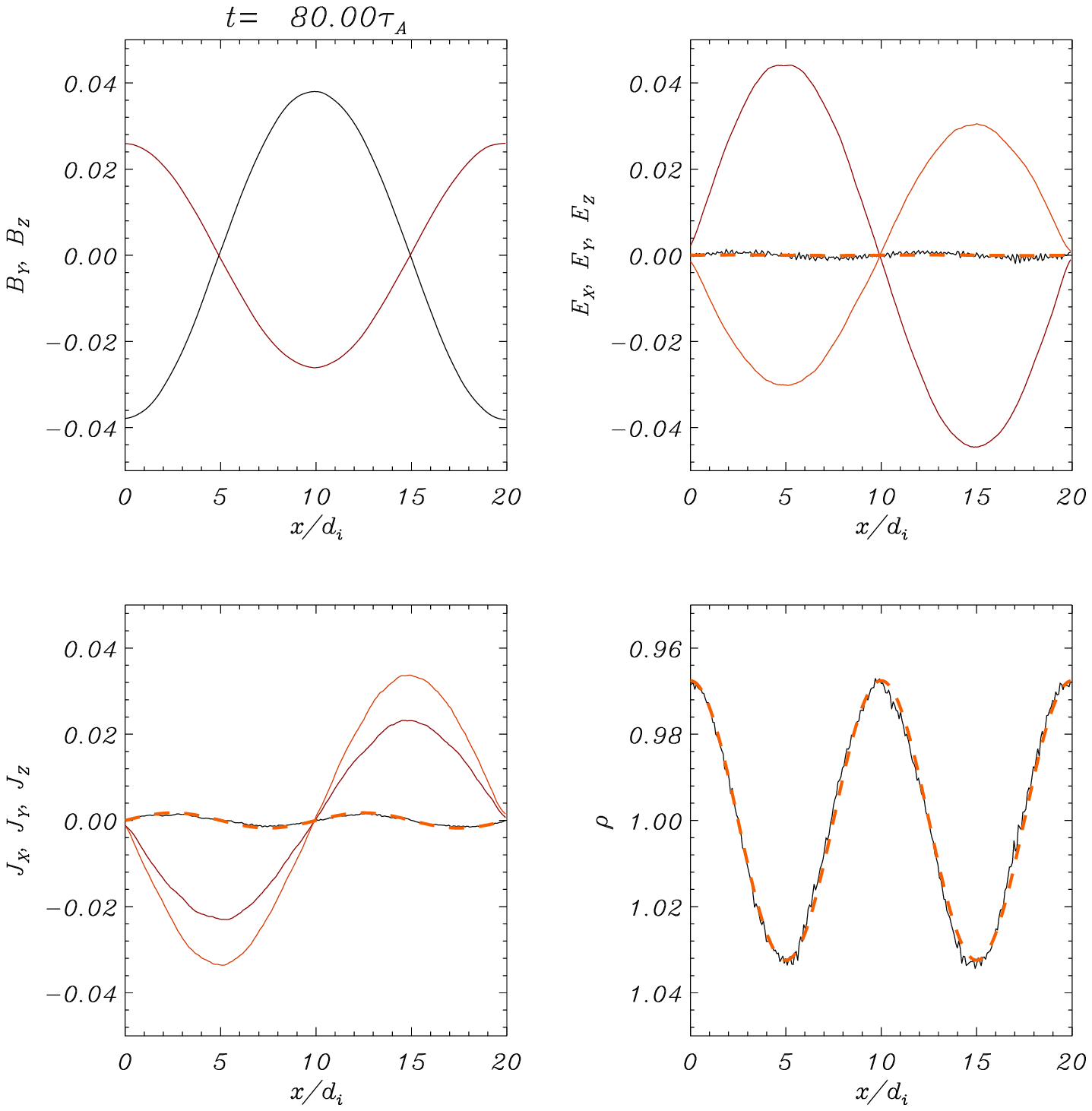}
} 
\caption{\small Top panel, macroscopic variables calculated from the hybrid simulations at $t=36.3\tau_{\rm A}$.  We have used  $\beta_{\rm e}=6\times 10^{-2}$ and $\beta_{\rm p}=10^{-3}$ and $k_0=2 \pi/L$. Bottom panel, same as the previous panel but at a later stage, $t=80\tau_{\rm A}$. The thick orange lines correspond to the analytical predictions for the parallel electric field, velocity (essentially $J_X$) and density given by Eqs.~(\ref{eparf}), (\ref{solvpar2}) and (\ref{solvpar2vp}), the match with the numerical results is quite significant. The temporal evolution is available as an online movie.
}\label{figionsnapnolandau}
\end{figure}

\subsection{Hybrid simulations results for the Alfv\'enic pump driver}

As in Sect.~\ref{secthybridsingle} we have performed hybrid simulations to have further confirmation of the previous analytical results based on the previous linear analysis. An example of a typical run is shown in Fig.~\ref{figionsnapnolandau}. The choice of the  ratio between the electron and proton temperatures makes the LD mechanism inefficient in this case and therefore suitable to be compared with the obtained analytical results. We chose to excite the system with a right-handed circularly polarised Alfv\'en wave with a wavelength that matches the size of our spatial domain (therefore periodic boundary conditions are applicable), i.e. $k_0=2 \pi/L$. In order to excite this type of wave we need to impose a certain profile for the velocity and magnetic field components at $t=0$, the details are provided in Appendix C, in particular we have to use the excitation given by Eqs.~(\ref{eqvstandingrl0})-(\ref{eqvstandingrlf}).  In the hybrid simulation the particle distribution function is initiated  self-consistently with the bulk velocity of the pump driver. The velocity and the magnetic field perturbations are related (see Eq.~(\ref{eqvpol})) through the fluid dispersion relation for the particular Alfv\'en driver pump wave (given by Eq.~(\ref{disperalf0})). There is no initial fluid speed in the parallel direction and no initial density perturbation either (since the Alfv\'enic wave is incompressible). In our case the initial distribution contains a delta function with drifts representing the transverse part and multiplying a Gaussian function with no longitudinal drift so that
\begin{align}       
f = \delta(v_\perp - u_{\perp 0})\, e^{-(v_x - u_{x 0})^2/{w_0}^2}
\end{align}
with $u_{x 0}=0$, see also Eqs.~(\ref{eq:10m}) and (\ref{eq:11}). This initial excitation is compatible, for example, with that of \citet{sonnerupsu1967}.

The simulation shows that we have excited the right-handed wave since its frequency agrees with the expected value, given by Eq.~(\ref{disperalfml}). We clearly find the driven ion-acoustic wave that is nonlinearly excited in the system and we are also able to reproduce its behaviour using the previous analytical results. For example, the density shows periodic spatial changes (cavities) with a wavenumber which is twice that of the Alfv\'enic driver ($2 k_0$). We have overplotted the analytical profiles for the parallel velocity and density fluctuation (Eqs.~(\ref{solvpar2}) and (\ref{solvpar2vp})) in Fig.~\ref{figionsnapnolandau} (see thick orange dashed lines) and we notice that the agreement with the simulation results is quite remarkable. In addition, the parallel electric field matches well with the formula given by Eq.~(\ref{eparf}). Similar results are obtained if instead of exciting the right-handed wave we perturb the system with a left-handed polarised wave, see Eqs.~(\ref{eqvstandingl0})-(\ref{eqvstandinglf}).

\begin{figure}
\centering{\includegraphics[width=8.cm]{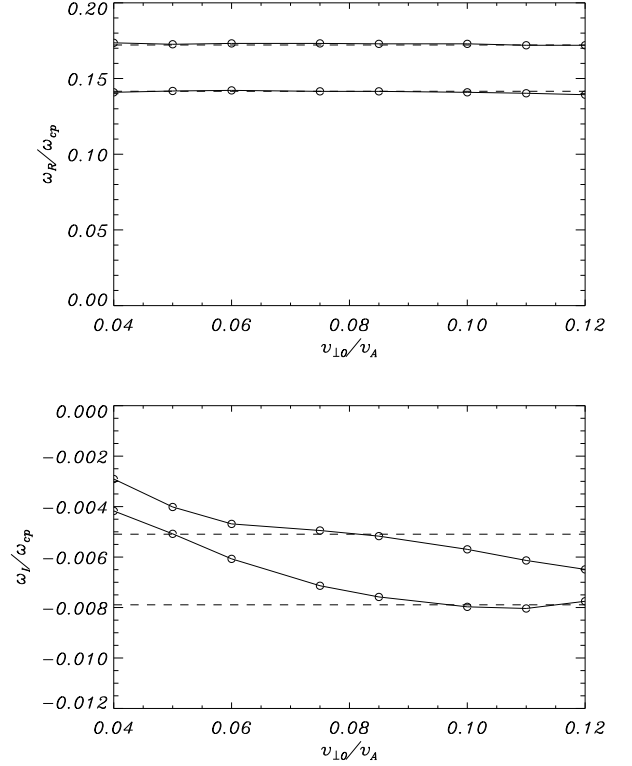}}
\caption{\small Real and imaginary parts of $\omega$ for a standing ion-acoustic  wave derived from hybrid simulations as a function of the amplitude of the Alfv\'enic pump. For the top curve, $\beta_{\rm e}=6\times 10^{-2}$ and $\beta_{\rm p}=10^{-2}$ while for the bottom curve, $\beta_{\rm e}=10^{-1}$ and $\beta_{\rm p}=1.25\times 10^{-2}$. In this simulation $k_0=2 \pi/L$. The attenuation is calculated from the Fourier analysis in space and the power of the mode $m'=2$ (or $m=4$). The horizontal dashed line corresponds to the expected real and imaginary frequency of the corresponding non-driven ion-acoustic mode attenuated by LD and calculated from the numerical solution of Eq.~(\ref{dispersionk}).
}\label{figiondamperiod}
\end{figure}

We now analyse how the driver affects the Landau damping of the excited wave. For comparison purposes we consider the same parameters in Sect.~\ref{secthybridsingle} where we studied the behaviour of an initially excited ion-acoustic wave. In the present simulations we find that the induced ion-acoustic shows an exponential damping in a similar way as it was found in Fig.~\ref{figiondamped}. Nevertheless, we find some dependence of the attenuation  on the amplitude of the Alfv\'enic pump and therefore on the amplitude of the ion-acoustic wave generated non-linearly. We already found this effect in the undriven case (see Sect.~\ref{secthybridsingle}). We have performed several simulations by changing the initial amplitude of the Alfv\'enic pump and have constructed  Fig.~\ref{figiondamperiod}. We find that the smaller the amplitude of the pump the weaker the attenuation by LD. Only when the initial velocity amplitude is around $0.1 v_{\rm A}$ the damping per period approaches the theoretical result of the non-driven ion-acoustic attenuated wave. However, we have to be cautious about this value of the pump since as we keep increasing its value more energy goes into the ion-acoustic wave (with an amplitude that is approximately given by Eq.~(\ref{constA})) which can eventually start developing shocks, as we found in Sect.~\ref{secthybridsingle}. In any case the agreement of the real part of omega with the expected value from the theoretical prediction is a clear indication that the Landau damping is taking place and that the process of attenuation if not significantly altered in the driven case. This agrees with the analytical justification that we have given in Sect.~\ref{secsecodord}.

\begin{figure}[!ht]
\centering{\includegraphics[width=9cm]{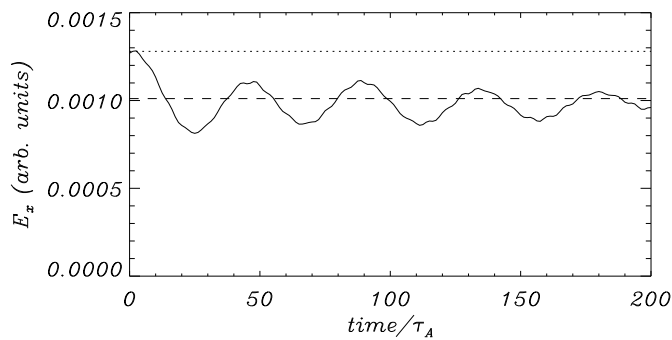}
} 
\caption{\small Parallel electric field amplitude associated to the mode $m'=2$ and calculated from the hybrid simulations.  The dotted line corresponds to the theoretical initial amplitude given by Eq.~(\ref{eparft0}) while the dashed line is the analytical estimation of the stationary value given by Eq.~(\ref{eparfstat}). The match with the numerical results is quite significant. In this simulation we have used $\beta_{\rm e}=6\times 10^{-2}$ and $\beta_{\rm p}=10^{-2}$ and the excitation corresponds to a $R-$mode.  
}\label{figefield}
\end{figure}

Finally, we investigate the generation of the parallel electric field in the simulations. We calculate the Fourier power associated to the $m'=2$ mode ($k=2 k_0$), which is the dominant wavenumber in the parallel electric field. The results from the hybrid calculations are shown in Fig.~\ref{figefield}. We find that the electric field starts at a certain value but then decreases and oscillates around a new value with the frequency of the driven ion-acoustic wave. This oscillation shows attenuation due to the LD process. The initial value of the parallel electric field agrees with the calculation given by Eq.~(\ref{eparft0}) valid a at $t=0$ (see dotted line). The asymptotic value matches also with the theoretical prediction for the stationary state given by Eq.~(\ref{eparfstat}) (see dashed line). This corroborates the validity of our theoretical analysis.

Finally, from the hybrid simulations we have investigated the possible presence of parametric instabilities. For most of the cases investigated in the present paper we have not found clear evidences of the presence parametric unstable modes. Only in some specific situations with very small LD damping and very long simulation times the effects of parametric instabilities can be identified. This confirms that the role of parametric instabilities is rather secondary in the present standing problem, as it has been demonstrated in the analysis given at the beginning of Sect.~\ref{sectdriven}.

\section{Application of the results}

We briefly describe a possible application of the results discussed so far. Slow modes, closely related to ion-acoustic modes, are often reported in the solar corona. In particular, observations indicate that standing waves slow modes are commonly excited in coronal loops \citep[see the review of][]{wangetal2021}. These modes are identified by Doppler and intensity fluctuations in hot coronal lines. A possible mechanism that can explain the presence of these modes, at least the first harmonic, is precisely a transverse Alfv\'enic oscillation that eventually drives nonlinearly ion-acoustic waves.  One of the most interesting features of the reported oscillations is that they are heavily damped with time, being the damping per period, $\tau_D/P$, of the order of one, meaning a severe attenuation. Several mechanism, including non-adiabatic effects such as thermal conduction, compressive viscosity,
and optically thin radiation have been proposed. Also nonlinearity, the cooling loop background, the wave-caused heating/cooling imbalance, plasma non-uniformities,
and other effects such as loop geometry, wave leakage through footpoints and in corona and magnetic effects (mode coupling and obliqueness) have been used to try to explain the fast attenuation. \citet{voitenko2005} concluded that the observed dissipation distances for propagating slow modes can alternatively be explained by phase mixing in its ideal regime, where the apparent damping is due to the spatial integration of the phase mixed amplitudes by the observation. 

The measurements of the difference in electron and ion temperature in the solar corona indicate in some cases that $T_{\rm e}\approx T_{\rm p}/3.5$ \citep[e.g.][]{boldyrevetal2020}. This value suggests that ion-acoustic waves suffer strong LD under coronal conditions and therefore this wave-particle interaction mechanism may explain, for example, the reported attenuation of standing slow modes in coronal loops. The damping per period due to Landau damping is essentially independent of the wavenumber, see Eqs.~(\ref{wrdis}) and (\ref{widis}), meaning that, although we have focused in the present work on very small spatial scales, in reality the mechanism operates on much larger spatial scales as well. Similar results concerning kinetic damping of small scale slow-wave modes and higher frequency waves were obtained by \citet{vinasetal2000} with regards to coronal heating and the formation of suprathermal tails in the coronal plasma. Therefore, we propose here that the attenuation of standing slow modes reported in the solar corona might be linked to the process of LD.
The expression for the damping per period using Eqs.~(\ref{wrdis}) and (\ref{widis}) is
\begin{align}\label{tdop}
    \tau_D/P=\frac{\sqrt{2}}{\pi^{3/2}}e^{\frac{1}{2}(\psi+3)}\frac{1}{\psi}\frac{1}{\sqrt{\psi+3}},
\end{align}
where
\begin{align}
    \psi=\gamma_{\rm e}\frac{\beta_{\rm e}}{\beta_{\rm p} }.
\end{align}
This expression is only applicable in the situation $\beta_{\rm p}\leq \beta_{\rm e}$, i.e., for weak attenuation. Equation ~(\ref{tdop}) can be used in the future for seismological purposes. In particular, it might be possible of infer the ratio $\beta_{\rm e}/\beta_{\rm p}$ based on the reported characteristics of the damped ion-acoustic modes.

The  effect of LD has been investigated, in the coronal context, for propagating waves, see \citet{ruanetal2016,ruanetal2016b}. These authors performed simulations of driven slow modes along the magnetic field by using a kinetic model that includes the effects of Landau damping and Coulomb collisions. They found that the obtained spatial scale of damping is similar to that resulting from thermal conduction in MHD. Future studies should address the case of the standing wave problem studied here, but including the effect of collisions which are not negligible in the lower solar corona and might have a significant effect on the attenuation by LD.

\section{Conclusions and discussion}\label{discussion}

Two cases of the evolution of ion-acoustic waves have been investigated. We have first studied the excitation of purely ion-acoustic standing modes using a multi-fluid approach. We have solved analytically the initial and boundary value problem incorporating kinetic effects by using a complex polytropic index.  We have shown that ion-acoustic standing waves attenuate by Landau damping, a result that is completely equivalent to the Landau attenuation for propagating ion-acoustic waves, extensively explored in the literature. The performed numerical hybrid simulations have shown however that the attenuation follows the predictions of the LD up to a certain degree. Too large or too small initial excitation amplitudes do not lead to the expected damping times due to the effects of nonlinearity and/or noise. In general we find that for the standing case, two symmetric beams develop in the distribution function since the standing wave can be viewed as the superposition of the two counter-propagating waves.

 The presence of two symmetric counter-streaming beams in the distribution function can have important consequences regarding the stability of the system. For example, it is well known in plasma physics, that counter-streaming beams can be the source of the oscillating two-stream instability that strongly affects the stability of such distributions and becomes the source for additional wave mode excitation. In the future, this aspect should be investigated in detail since two stream instabilities may develop and affect the dynamics of the configuration. In essence when two streams move through each other one wavelength in one cycle of the plasma frequency, a density perturbation on one stream is reinforced by the forces due to bunching of particles in the other stream and vice versa \citep[e.g.][]{dawson1962,birdsalllangdon1991}.

The second aspect investigated in this paper considers the case when the Alfv\'enic pump is present, driving the ion-acoustic waves. For this case we have shown that standing circularly polarised waves have the same dispersion relation as propagating waves and we have derived the appropriate polarisation conditions for their excitation. We further have demonstrated that, parametric instabilities are secondary in the standing problem, contrary to the situation studied by \citet{chianoliverira1994, chianoliverira1996, oliverirachian1996}, and to the purely propagating case. The Alfv\'enic pump excites nonlinearly, via ponderomotive forces, the second harmonic of the standing ion-acoustic wave, i.e., with $k=2 k_0$, where $k_0$ is the wavenumber of the standing pump. The frequency of the ion-acoustic wave is $\omega=2 k_0 C_{\rm s}$, in the absence of Landau attenuation. When Landau damping is relevant it also affects the real part of the frequency. These ion-acoustic waves are damped with time depending on the ratio $T_{\rm e}/T_{\rm p}$. The hybrid simulations have shown that the attenuation follows the standard linear predictions in the non-driven case up to a certain degree. The attenuation is weaker when the amplitude of the driver is low.

We have derived analytical expressions for the density and electric fluctuations induced quadratically by the pump. Left and right-handed modes produce exactly the same density fluctuations but we have demonstrated that the $L-$mode has the largest parallel electric field fluctuation. These results provide a possible interpretation of measurements of electric fields parallel to the magnetic field observed, for example, in the auroral ionosphere. Although several mechanisms can explain this reported electric field, \citet{isofman2004,isofman2011} have proposed that precisely standing Alfv\'en waves can naturally generate the parallel electric field. The analytic expressions for the generated electric fields could be used in the future to compare with the observations, although the first impression is that this electric field is rather small in general. On a different context, \citet{mottez2015J} has investigated the generation of electric fields due to the interaction of counter propagating Alfv\'en waves. Nevertheless, the generated electric field in our case is purely due to the ponderomotive forces, while the electric field described by the previous author has a completely different origin. 

We have proposed that standing slow modes reported in coronal loops can be interpreted in terms of ion-acoustic waves that attenuate by Landau damping. The ratio $T_{\rm e}/T_{\rm p}$ plays the relevant role in this problem, and may be it can be indirectly inferred from the damping per period of slow modes reported from the observations. However, the low corona is known to be weakly collisional and this can be a competing effect that can have a significant influence on the process of LD. The efficiency of the attenuation is altered by the thermalisation of the non-thermal tail in the velocity distribution by Coulomb collisions. In the propagating case, beam flows are produced by Landau resonance but they are destroyed by Coulomb friction. From the theoretical point of view the inclusion of  collisional effects is through the Vlasov equation by including a Fokker-Planck collisional term, although there are other alternatives. This problem should be investigated in the near future with potential applications to partially ionised plasmas also.

The reasons to neglect electron Landau damping are that the focus of this paper is to study low frequency waves (i.e. waves below the ion-cyclotron and ion-plasma frequencies) for which electron Landau damping has little or no-contribution since this effect is more important for high frequency waves (i.e. like  whistler, electron-acoustic, and Langmuir waves). Furthermore, because the ion and electron temperatures are similar, the electron thermal velocity (i.e. the width) of the electron velocity distribution function is about 43 times larger than the ion thermal velocity. This indicates that the electron LD is much smaller than that of ion-Landau damping for resonances near the ion-thermal speed since LD is proportional to the slope of the velocity distribution function. Only ion-Landau damping is included in this work. Also, the use of the hybrid simulation neglects by construction electron LD since electrons are treated as fluid. Additionally, there are various reasons to consider only zero ion perpendicular temperature in our study. This is due to the fact that a finite ion-perpendicular temperature has no effect on Landau damping and only contributes to ion-cyclotron damping when finite ion-gyroradius effects are included. In principle, finite perpendicular temperature effects can be included, however the mathematical calculations become more tediously complex since in this case one needs to introduce not only a single cyclotron frequency but also an infinite series of cyclotron-harmonic that will require a truncation. And more important the effect of finite perpendicular temperature will not contribute to LD at all, only to cyclotron damping. Since in our study we are only considering parallel propagation, this fact further eliminates the effect of finite ion-perpendicular temperatures. The hybrid simulations do include finite-temperature effects such as cyclotron-damping but notice that they do not change the Landau damping which is a longitudinal effect.

The results presented here will be used to consider more realistic situations. In particular, it is interesting to introduce different ion populations to study their effects. For example, two counter-propagating light beams together with  a heavy stationary population. This could represent a loop with weak flows penetrating thought the footpoints. The stability of this system can be then analysed using kinetic theory with both analytical methods and hybrid simulations. Interestingly, different types of instabilities can appear in the system. Along the same line, we have focused on initial excitations that correspond to pure standing waves, meaning that the two counter propagating waves have exactly the same amplitude and wavelength but propagating in opposite directions. It would be interesting to analyse the effect of two slightly unbalanced counter propagating waves, i.e. when the symmetry is broken. This can significantly change the dynamics of the system, lead to the appearance of more clear beams and possibly to the development of stronger parametric instabilities, which are weak in the perfectly symmetric case.


\begin{acknowledgements}
J.T. acknowledges the support from grant AYA2017-85465-P
(MINECO/AEI/FEDER, UE), to the Conselleria d'Innovaci\'o, Recerca i Turisme del
Govern Balear, and also to IAC$^3$. A. F.- Vinas would like to thank the Universitat de les Ilhes Balears for their assistance as a visiting professor and to the Catholic University/IACS and NASA-GSFC for their support during the development of this work. J.A. acknowledges the support from FONDECYT grant 1161700. The authors thank Prof. Marcel Goossens for useful suggestions that helped to improve the paper. We thank the anonymous referee for their useful comments that have helped to improve this manuscript.
\end{acknowledgements}

\bibliographystyle{aa}
\bibliography{jaume}

\begin{thebibliography}{59}
\expandafter\ifx\csname natexlab\endcsname\relax\def\natexlab#1{#1}\fi

\bibitem[{{Araneda}(1998)}]{araneda1998}
{Araneda}, J.~A. 1998, Physica Scripta Volume T, 75, 164

\bibitem[{{Araneda} {et~al.}(2008){Araneda}, {Marsch}, \& {F.
  -Vi{\~n}as}}]{aranedaetal2008}
{Araneda}, J.~A., {Marsch}, E., \& {F. -Vi{\~n}as}, A. 2008, \prl, 100, 125003

\bibitem[{{Araneda} {et~al.}(2007){Araneda}, {Marsch}, \&
  {Vi{\~n}as}}]{aranedaetal2007}
{Araneda}, J.~A., {Marsch}, E., \& {Vi{\~n}as}, A.~F. 2007, Journal of
  Geophysical Research (Space Physics), 112, A04104

\bibitem[{{Araneda} {et~al.}(2002){Araneda}, {Vi{\~n}as}, \&
  {Astudillo}}]{aranedaetal2002}
{Araneda}, J.~A., {Vi{\~n}as}, A.~F., \& {Astudillo}, H.~F. 2002, Journal of
  Geophysical Research (Space Physics), 107, 1453

\bibitem[{{Aschwanden} {et~al.}(2002){Aschwanden}, {de Pontieu}, {Schrijver},
  \& {Title}}]{aschetal02}
{Aschwanden}, M.~J., {de Pontieu}, B., {Schrijver}, C.~J., \& {Title}, A.~M.
  2002, \solphys, 206, 99

\bibitem[{{Ballester} {et~al.}(2020){Ballester}, {Soler}, {Terradas}, \&
  {Carbonell}}]{ballesteretal2020}
{Ballester}, J.~L., {Soler}, R., {Terradas}, J., \& {Carbonell}, M. 2020, \aap,
  641, A48

\bibitem[{{Birdsall} \& {Langdon}(1991)}]{birdsalllangdon1991}
{Birdsall}, C.~K. \& {Langdon}, A.~B. 1991, {Plasma Physics via Computer
  Simulation}

\bibitem[{{Boldyrev} {et~al.}(2020){Boldyrev}, {Forest}, \&
  {Egedal}}]{boldyrevetal2020}
{Boldyrev}, S., {Forest}, C., \& {Egedal}, J. 2020, arXiv e-prints,
  arXiv:2001.05125

\bibitem[{{Chian} \& {Oliveira}(1994)}]{chianoliverira1994}
{Chian}, A.~C.~L. \& {Oliveira}, L.~P.~L. 1994, \aap, 286, L1

\bibitem[{{Chian} \& {Oliveira}(1996)}]{chianoliverira1996}
{Chian}, A.~C.~L. \& {Oliveira}, L.~P.~L. 1996, \aap, 309, 673

\bibitem[{{Cramer}(2001)}]{cramer2001}
{Cramer}, N.~F. 2001, {The Physics of Alfv{\'e}n Waves}

\bibitem[{Cross(1988)}]{cross1988}
Cross, R. 1988, An Introduction to Alfven Waves, (Taylor \& Francis)

\bibitem[{{Cummings} {et~al.}(1969){Cummings}, {O'Sullivan}, \&
  {Coleman}}]{cummingsetal1969}
{Cummings}, W.~D., {O'Sullivan}, R.~J., \& {Coleman}, P.~J., J. 1969, Journal
  of Geophysical Research (Space Physics), 74, 778

\bibitem[{{Danielson} {et~al.}(2002){Danielson}, {Anderegg}, {Shiga}, {Rigg},
  \& {Driscoll}}]{danielsonetal2002}
{Danielson}, J.~R., {Anderegg}, F., {Shiga}, N., {Rigg}, K.~M., \& {Driscoll},
  C.~F. 2002, in APS Meeting Abstracts, Vol.~44, APS Division of Plasma Physics
  Meeting Abstracts, KO1.006

\bibitem[{{Dawson}(1962)}]{dawson1962}
{Dawson}, J. 1962, Physics of Fluids, 5, 445

\bibitem[{{Derby}(1978)}]{derby1978}
{Derby}, N.~F., J. 1978, \apj, 224, 1013

\bibitem[{{Goldstein}(1978)}]{goldstein1978}
{Goldstein}, M.~L. 1978, \apj, 219, 700

\bibitem[{{Gomberoff} \& {Elgueta}(1991)}]{gomberoffelgueta1991}
{Gomberoff}, L. \& {Elgueta}, R. 1991, \jgr, 96, 9801

\bibitem[{{Hammett} \& {Perkins}(1990)}]{hammetperkins1990}
{Hammett}, G.~W. \& {Perkins}, F.~W. 1990, \prl, 64, 3019

\bibitem[{{Hasegawa} \& {Uberoi}(1982)}]{hasegawauberoi1982}
{Hasegawa}, A. \& {Uberoi}, C. 1982, {The Alfv{\'e}n wave.}

\bibitem[{{Hollweg}(1971)}]{hollweg71}
{Hollweg}, J.~V. 1971, \jgr, 76, 5155

\bibitem[{{Hollweg}(1994)}]{hollweg1994}
{Hollweg}, J.~V. 1994, \jgr, 99, 23431

\bibitem[{{Horowitz} {et~al.}(1989){Horowitz}, {Shumaker}, \&
  {Anderson}}]{horowitz1989}
{Horowitz}, E.~J., {Shumaker}, D.~E., \& {Anderson}, D.~V. 1989, Journal of
  Computational Physics, 84, 279

\bibitem[{{Israelevich} \& {Ofman}(2004)}]{isofman2004}
{Israelevich}, P. \& {Ofman}, L. 2004, Annales Geophysicae, 22, 2797

\bibitem[{{Israelevich} \& {Ofman}(2011)}]{isofman2011}
{Israelevich}, P.~L. \& {Ofman}, L. 2011, Advances in Space Research, 48, 25

\bibitem[{{Jayanti} \& {Hollweg}(1993)}]{jayantihollweg1993}
{Jayanti}, V. \& {Hollweg}, J.~V. 1993, \jgr, 98, 13247

\bibitem[{{Khurana}(1993)}]{khurana1993}
{Khurana}, K.~K. 1993, Annales Geophysicae, 11, 973

\bibitem[{{Lancellotti} \& {Dorning}(1998)}]{lancelottidoring1998}
{Lancellotti}, C. \& {Dorning}, J.~J. 1998, \prl, 81, 5137

\bibitem[{{Lancellotti} \& {Dorning}(2009)}]{lancelottidoring2009}
{Lancellotti}, C. \& {Dorning}, J.~J. 2009, Transport Theory and Statistical
  Physics, 38, 1

\bibitem[{{Lundin} \& {Guglielmi}(2006)}]{lundinguglielmi2006}
{Lundin}, R. \& {Guglielmi}, A. 2006, \ssr, 127, 1

\bibitem[{{Maneva} {et~al.}(2013){Maneva}, {Vi{\~n}as}, \&
  {Ofman}}]{manevaetal2013}
{Maneva}, Y.~G., {Vi{\~n}as}, A.~F., \& {Ofman}, L. 2013, Journal of
  Geophysical Research (Space Physics), 118, 2842

\bibitem[{{Manners} {et~al.}(2018){Manners}, {Masters}, \&
  {Yates}}]{manneretal2018}
{Manners}, H., {Masters}, A., \& {Yates}, J.~N. 2018, Geophysical Research
  Letters, 45, 8746

\bibitem[{{Mart{\'\i}nez-G{\'o}mez} {et~al.}(2018){Mart{\'\i}nez-G{\'o}mez},
  {Soler}, \& {Terradas}}]{davidetal2018}
{Mart{\'\i}nez-G{\'o}mez}, D., {Soler}, R., \& {Terradas}, J. 2018, \apj, 856,
  16

\bibitem[{{Mjolhus} \& {Wyller}(1988)}]{mjlhuswyller1988}
{Mjolhus}, E. \& {Wyller}, J. 1988, Journal of Plasma Physics, 40, 299

\bibitem[{{Mottez}(2015)}]{mottez2015J}
{Mottez}, F. 2015, Journal of Plasma Physics, 81, 325810104

\bibitem[{{Moya} {et~al.}(2012){Moya}, {Vi{\~n}as}, {Mu{\~n}oz}, \&
  {Valdivia}}]{moyaetal2012}
{Moya}, P.~S., {Vi{\~n}as}, A.~F., {Mu{\~n}oz}, V., \& {Valdivia}, J.~A. 2012,
  Annales Geophysicae, 30, 1361

\bibitem[{{Nakariakov} {et~al.}(2021){Nakariakov}, {Anfinogentov}, {Antolin},
  {Jain}, {Kolotkov}, {Kupriyanova}, {Li}, {Magyar}, {Nistic{\`o}}, {Pascoe},
  {Srivastava}, {Terradas}, {Vasheghani Farahani}, {Verth}, {Yuan}, \&
  {Zimovets}}]{nakaetal2021}
{Nakariakov}, V.~M., {Anfinogentov}, S.~A., {Antolin}, P., {et~al.} 2021, Space
  Sience Reviews, 217, 73

\bibitem[{{Nakariakov} {et~al.}(1999){Nakariakov}, {Ofman}, {Deluca},
  {Roberts}, \& {Davila}}]{nakaetal99}
{Nakariakov}, V.~M., {Ofman}, L., {Deluca}, E.~E., {Roberts}, B., \& {Davila},
  J.~M. 1999, Science, 285, 862

\bibitem[{{Oliveira} \& {Chian}(1996)}]{oliverirachian1996}
{Oliveira}, L.~P.~L. \& {Chian}, A.~C.~L. 1996, Journal of Plasma Physics, 56,
  251

\bibitem[{{O'Neil}(1965)}]{oneil1965}
{O'Neil}, T. 1965, Physics of Fluids, 8, 2255

\bibitem[{{Podesta} \& {Gary}(2011)}]{podestagary2011}
{Podesta}, J.~J. \& {Gary}, S.~P. 2011, ApJ, 742, 41

\bibitem[{{Rankin} {et~al.}(1994){Rankin}, {Frycz}, {Tikhonchuk}, \&
  {Samson}}]{rankinetal94}
{Rankin}, R., {Frycz}, P., {Tikhonchuk}, V.~T., \& {Samson}, J.~C. 1994, \jgr,
  99, 21291

\bibitem[{{Ruan} {et~al.}(2016{\natexlab{a}}){Ruan}, {He}, {Zhang}, {Vocks},
  {Marsch}, {Tu}, {Peter}, \& {Wang}}]{ruanetal2016}
{Ruan}, W., {He}, J., {Zhang}, L., {et~al.} 2016{\natexlab{a}}, \apj, 825, 58

\bibitem[{{Ruan} {et~al.}(2016{\natexlab{b}}){Ruan}, {He}, {Zhang}, {Vocks},
  {Marsch}, {Tu}, {Peter}, \& {Wang}}]{ruanetal2016b}
{Ruan}, W., {He}, J., {Zhang}, L., {et~al.} 2016{\natexlab{b}}, in American
  Institute of Physics Conference Series, Vol. 1720, Solar Wind 14, 020004

\bibitem[{{Scargle}(1982)}]{scargle1982}
{Scargle}, J.~D. 1982, \apj, 263, 835

\bibitem[{{Schrijver} {et~al.}(2002){Schrijver}, {Aschwanden}, \&
  {Title}}]{schrijetal02}
{Schrijver}, C.~J., {Aschwanden}, M.~J., \& {Title}, A.~M. 2002, \solphys, 206,
  69

\bibitem[{{Sonnerup} \& {Su}(1967)}]{sonnerupsu1967}
{Sonnerup}, B.~U.~{\"O}. \& {Su}, S.-Y. 1967, Physics of Fluids, 10, 462

\bibitem[{{Spangler}(1989)}]{spangler1989}
{Spangler}, S.~R. 1989, Physics of Fluids B, 1, 1738

\bibitem[{{Swanson}(2003)}]{swanson2003}
{Swanson}, D.~G. 2003, Plasma Physics and Controlled Fusion, 45, 1069

\bibitem[{{Terasawa} {et~al.}(1986){Terasawa}, {Hoshino}, {Sakai}, \&
  {Hada}}]{terasawa1986}
{Terasawa}, T., {Hoshino}, M., {Sakai}, J.~I., \& {Hada}, T. 1986, \jgr, 91,
  4171

\bibitem[{{Tikhonchuk} {et~al.}(1995){Tikhonchuk}, {Rankin}, {Frycz}, \&
  {Samson}}]{tikhoetal1995}
{Tikhonchuk}, V.~T., {Rankin}, R., {Frycz}, P., \& {Samson}, J.~C. 1995,
  Physics of Plasmas, 2, 501

\bibitem[{{Vasquez}(1995)}]{vasquez1995}
{Vasquez}, B.~J. 1995, \jgr, 100, 1779

\bibitem[{{Vi{\~n}as} {et~al.}(2000){Vi{\~n}as}, {Wong}, \&
  {Klimas}}]{vinasetal2000}
{Vi{\~n}as}, A.~F., {Wong}, H.~K., \& {Klimas}, A.~J. 2000, \apj, 528, 509

\bibitem[{{Voitenko} {et~al.}(2005){Voitenko}, {Andries}, {Copil}, \&
  {Goossens}}]{voitenko2005}
{Voitenko}, Y., {Andries}, J., {Copil}, P.~D., \& {Goossens}, M. 2005, \aap,
  437, L47

\bibitem[{{Wang} {et~al.}(2021){Wang}, {Ofman}, {Yuan}, {Reale}, {Kolotkov}, \&
  {Srivastava}}]{wangetal2021}
{Wang}, T., {Ofman}, L., {Yuan}, D., {et~al.} 2021, \ssr, 217, 34

\bibitem[{{Winske} \& {Leroy}(1985)}]{winskeleroy1985}
{Winske}, D. \& {Leroy}, M.~M. 1985, in Computer Simulation of Space Plasmas,
  255--278

\bibitem[{{Winske} \& {Omidi}(1996)}]{winskeomidi1996}
{Winske}, D. \& {Omidi}, N. 1996, \jgr, 101, 17287

\bibitem[{{Wong} \& {Goldstein}(1986)}]{wonggolstein1986}
{Wong}, H.~K. \& {Goldstein}, M.~L. 1986, \jgr, 91, 5617

\bibitem[{{Xie} {et~al.}(2004){Xie}, {Ofman}, \& {Vi{\~n}as}}]{xieetal2004}
{Xie}, H., {Ofman}, L., \& {Vi{\~n}as}, A. 2004, Journal of Geophysical
  Research (Space Physics), 109, A08103

\end{thebibliography}

\newpage

\begin{appendix}
\section{Incorporating kinetic effects in the fluid description}
The method was presented \citet{araneda1998,aranedaetal2007} for propagating waves but here we extend it to the case of standing waves. Kinetic effects in the fluid equations emerge through the Vlasov equation,
\begin{align}\label{vlasov}
		\frac{\partial F_{s}(x,{\bf v},t)}{\partial t}&+{\bf v}\cdot \nabla F_{s}(x,{\bf v},t)\nonumber\\
		&+\frac{q_{s}}{m_{s}}\left({\bf {E}}(x,t) + \frac{{\bf v} \times {\bf {B}}(x,t)}{c}\right) \frac{\partial F_{s}(x,{\bf v},t)}{\partial {\bf v}}=0,\nonumber\\
\end{align}
where $F_{s}(x,{\bf v},t)$ represents the distribution function of species $s$ and ${\bf v}$ is the velocity vector that corresponds to the particle distribution velocities. Do not confuse this velocity with the fluid velocity, {\bf u}, introduced in the main text. 

We linearise Eq.~(\ref{vlasov}) with
\begin{align}
F_{s}(x,{\bf v},t) &= F_{0s}(x,{\bf v},t)+\delta f_{s}(x,{\bf v},t),\label{eq:10}\\
F_{0s}(x,{\bf v},t) &= \delta \left({\bf v}_\bot-{\bf u_{\perp}}_{0s}(x,t)\right)f_{0s}(v_{x}),\label{eq:10m}
\end{align}
where we have assumed that the fluctuation of the velocity distribution function is $ \delta f_{s}(x,v_{x},t) $. In Eq.~(\ref{eq:10m}) we have assumed that the perpendicular temperature in the zeroth order is zero as represented by the Dirac delta-function in the transverse velocity direction which includes only the effect of the transverse wave velocity drift. The zeroth order parallel velocity distribution $ f_{0s}(v_{x}) $ can be taken, with all generality, as a finite temperature drifting Maxwellian distribution given by
  \begin{equation}\label{eq:11}
 f_{0s}(v_{x})=\frac{n_{0s}}{\sqrt{\pi}w_{0s}} e^{-\frac{(v_{x}-u_{x0s})^2}{w_{0s}^2}},
 \end{equation}
 \noindent  where $n_{0s}$ is the equilibrium ion density and $ w_{0s}=\sqrt{2\kappa_{B} T_{0s}/m_{s}}$ is the thermal velocity and $u_{x0s}$ is the velocity drift. The first-order linearised expression for $ \delta f_{s}(x,v_{x},t) $ can be immediately determined as
 \begin{align}\label{eq:12}
  		\frac{\partial \delta f_{s}(x,v_{x},t)}{\partial t}&+v_{x}\frac{\partial \delta f_{s}(x,v_{x},t)}{\partial x}
 +\frac{q_{s}}{m_{s}}\Bigl( \delta E_{x}(x,t)+ \nonumber \\&\frac{\delta u_{ys}B_{z}-\delta u_{zs}B_{y}+{u}_{ys}\delta B_{z}-{u}_{zs}\delta B_{y}}{c}\Bigr) \frac{\partial f_{0s}(v_{x})}{\partial v_{x}}=0.\nonumber \\
 \end{align}
 We assume that the fluctuations vary as $e^{i(k x-\omega t)}$ with a complex frequency $ \omega $ and a real wave vector $k$. We are therefore considering propagating waves for the moment, but as mentioned earlier the goal is to find a solution for standing waves. 
 Following this linearising procedure in Eq.~(\ref{eq:12}) yields to
\begin{eqnarray}\label{eq:13}
\delta f_{s}(x,v_{x},t) = -i\frac{q_{s}}{m_{s}} \delta G_{s}(x,t) \frac{1}{\omega - k v_x} \frac{\partial f_{0s}(v_{x})}{{\partial v_{x}}},
\end{eqnarray}
\noindent where we have defined 
\begin{eqnarray}\label{eq:13m}
\delta G_{s}(x,t)=\delta E_{x} + \frac{\delta {u}_{ys}B_{z}-\delta {u}_{zs}B_{y}+{u}_{ys}\delta B_{z}-{u}_{zs}\delta B_{y}}{c},  \end{eqnarray}
\noindent that represents a longitudinal force, and it is dependent in $ (x,t) $ but independent of the particle velocities. At this point is unnecessary to know the exact form of this force, thus we will focus on the resonant wave-particle interaction term in Eq.~(\ref{eq:13}). Note that for purely ion-acoustic waves propagating with parallel propagation, the magnetic terms are zero in Eq.~(\ref{eq:13m}) and only the $x-$component of the electric field survives.

We now calculate the moment density fluctuation $ \delta n_{s} $ of the plasma from the linearised velocity distribution fluctuation $ \delta f_{s}(x,v_{x},t) $ as:
\begin{equation}\label{eq:14}
\delta n_{s}(x,t) =\int_{-\infty}^{+\infty} \delta f_{s}(x,v_{x},t) \,dv_{x}.
\end{equation}
\noindent By substitution of Eq.~(\ref{eq:11}) and Eq.~(\ref{eq:13}) into Eq.~(\ref{eq:14}) and performing the integrals (see the details in Appendix B) we obtain
\begin{equation}\label{eq:15}
\delta n_{s}(x,t) =i n_{0s}\frac{q_{s}}{m_{s}} \frac{\delta G_{s}(x,t)}{k w_{0s}^2} Z'(\xi_{s}),
\end{equation}
which is written in terms of the derivative of the Zeta function, $Z$ and $ \xi_{s}=\frac{\omega - k u_{x0s}}{k\, w_{0s}}$. We now associate the density fluctuation derived in Eq.~(\ref{eq:15}) to the pressure fluctuation via the usual polytrope equation
\begin{equation}\label{eq:16}
\frac{\delta p_{s}}{p_{0s}}=\gamma \frac{\delta n_{s}}{n_{0s}},
\end{equation}
where $ \gamma $ is the polytrope coefficient that is not constant anymore, but depends on the complex frequency, the wave vector and the background equilibrium parameters of the plasma. To obtain the proper representation of the polytrope $ \gamma $ we need to use the linearised continuity equation and the linearised longitudinal fluid momentum equation for species $s$ given by:
\begin{align}\label{eq:17}
\frac{\partial \delta n_{s}}{\partial t}&+\frac{\partial (\delta n_{s}{u}_{0xs}+n_{0s}\delta {u}_{xs})}{\partial x}=0,\\
\frac{\partial \delta {u}_{xs}}{\partial t}+&{u}_{0xs}\frac{\partial \delta {u}_{xs}}{\partial x}=\nonumber \\ &\frac{q_{s}}{m_{s}}\left(\delta E_{x} + \frac{\delta {u}_{ys}B_{z}-\delta {u}_{zs}B_{y}+{u}_{ys}\delta B_{z}-{u}_{zs}\delta B_{y}}{c}\right)\nonumber \\
&-\frac{1}{m_{s}n_{0s}}\frac{\partial \delta p_{s}}{\partial x}. \label{eq:18}
\end{align}

\noindent Upon linearisation of Eq.~(\ref{eq:17}) and Eq.~(\ref{eq:18}) we obtain
\begin{align}
(\omega - k\,{u}_{0xs})\, \delta n_{s}&=k n_{0s}\, \delta {u}_{xs},\label{eq:19} \\
-i(\omega - k\, {u}_{0xs})\, \delta {u}_{xs}&=\frac{q_{s}}{m_{s}} \delta G_{s}(x,t) - i\frac{k \delta p_{s}}{m_{s}n_{0s}}.\label{eq:20}
\end{align}

\noindent From Eq.~(\ref{eq:19}) we get an expression for $\delta {u}_{xs}$ which is substituted in Eq.~(\ref{eq:20}). Combined with Eq.~(\ref{eq:15}) it leads to the complex kinetic polytropic relationship
\begin{equation}\label{eq:21}
\frac{\delta p_{s}}{p_{0s}}=2\left(\xi_{s}^2-\frac{1}{Z'(\xi_{s})}\right) \frac{\delta n_{s}}{n_{0s}}.
\end{equation}
where $Z'$ is the derivative of the plasma dispersion function. We have also used that $p_{0s}=n_{0s} k_{\rm B} T_{0s}$.

Therefore, comparing Eq.~(\ref{eq:16}) with Eq.~(\ref{eq:21}) we recognise the complex kinetic polytrope coefficient as
\begin{equation}\label{eq:22}
\gamma = 2\left(\xi_{s}^2-\frac{1}{Z'(\xi_{s})}\right).
\end{equation}

\noindent Equations (\ref{eq:21}) and (\ref{eq:22})  clearly show that density and pressure fluctuations are not necessarily in phase in contrast with the phase-locking that results from the constant value obtained in the fluid treatment. This result indicates that the kinetic polytropic coefficient depends on the dispersive properties of the plasma. Since this coefficient is in the Fourier-domain, its inverse transform (e.g., via a superposition of the wave modes that corresponds to multiple spatial scales of the system) of Eq.~(\ref{eq:22}) implies an intricate nonlocal real-space relationship between pressure and density of the system. On the contrary, in the fluid treatment the polytrope is assumed to be constant and the pressure and density relationship is local. Therefore, the dispersive property of the polytrope coefficient introduces nonlocal effects which ultimate control the transfer of energy between finite amplitude transverse and longitudinal fluctuations \citep{mjlhuswyller1988,spangler1989,hammetperkins1990}. As a result, this can explain to certain extent, why it is difficult to obtain reasonable values of the polytropic coefficient when comparing linear fluid theory with kinetic simulations \citep{vasquez1995}. We should emphasised, however, that we have included kinetics effects only in the parallel direction, but we kept the assumption for drifting Maxwellian distributions. This means that we keep only the collisionless damping effects of Landau damping on the growth rates of ion-acoustic wave, but all other kinetic cyclotron resonance effects (due to the simplified zeroth transverse temperature effects and finite gyro-radius effects) are completely neglected.

Let us now repeat the same analysis but for a propagating wave in the opposite direction, i.e., now the wave propagates to the left, meaning that we are assuming that the wavenumber is $-k$. Following the same procedure we now have that the fluctuation in the perturbed distribution function is simply,
\begin{eqnarray}\label{eq:13left}
\delta f_{s}(x,v_{x},t) = -i\frac{q_{s}}{m_{s}} \delta G_{s}(x,t) \frac{1}{\omega + k  v_x} \frac{\partial f_{0s}(v_{x})}{{\partial v_{x}}},
\end{eqnarray}
but note the change in the denominator with respect the equivalent equation for the right propagating wave in Eq.~(\ref{eq:13}). 

Calculating the density fluctuation $ \delta n_{s} $ of the plasma from the linearised velocity distribution fluctuation $ \delta f_{s}(x,v_{x},t) $ as before we obtain now
\begin{equation}\label{eq:15left}
\delta n_{s}(x,t) =-i n_{0s}\frac{q_{s}}{m_{s}} \frac{\delta G_{s}(x,t)}{k w_{0s}^2} Z'(\xi_{s}).
\end{equation}
where we use the definition for $\xi_s$ as in the previous case ($\xi_{s}=\frac{\omega}{ k w_{0s}}$, where for simplicity we have eliminated the drift). Details about this derivation are found in Appendix B. Note the change in the global sign in comparison with the result for the wave propagating to the right, Eq.~(\ref{eq:15}). The calculation of the  complex kinetic polytropic relationship leads now to
\begin{equation}\label{eq:21left}
\frac{\delta p_{s}}{p_{0s}}=2\left(\xi_{s}^2-\frac{1}{Z'(\xi_{s})}\right) \frac{\delta n_{s}}{n_{0s}},
\end{equation}
and therefore the complex kinetic polytrope coefficient is
\begin{equation}\label{eq:22left}
\gamma = 2\left(\xi_{s}^2-\frac{1}{Z'(\xi_{s})}\right).
\end{equation}
It is clear that from the physical point of view it should lead to exactly the same damping times as for the right propagating wave when there is no Alfv\'enic pump. And this agrees with the value of 
 $\gamma$ that we have obtained, it is exactly the same as in Eq.~(\ref{eq:22}) for the wave propagating to the right.

 These results are for propagating waves but we are interested on standing waves. The previous derivation leads us to think that in fact the full standing solution for the distribution function should be of the form
\begin{equation}\label{totaldistrib}
\delta f_{s}(x,v_{x},t)=f_+(v_x)\, e^{i(k x-\omega t)}+f_-(v_x)\, e^{i(-k x -\omega t)},
\end{equation}
where $f_+(v_x)$ and $f_-(v_x)$ are not equal, according to what we have shown before. Introducing this fluctuation in the Vlasov equation now we obtain the following
\begin{align}\label{vlasovlinstand}
-i\omega e^{-i \omega t}\left(f_+ e^{ikx}+f_- e^{-ikx}\right)&+v_x ik e^{-i \omega t} \left(f_+ e^{ikx}-f_- e^{-ikx}\right)\nonumber \\ &+\frac{q_s}{m_s}\delta G(x,t)\frac{\partial f_{0s}(v_{x})}{\partial v_{x}}=0 .
\end{align}
To have a proper solution, $\delta G(x,t)$ must be of the form
\begin{eqnarray}
\delta G(x,t)= e^{-i \omega t}\left(G_+e^{ikx} + G_-e^{-ikx} \right),
\end{eqnarray}
in this case the temporal phases cancel and to have a solution the spatial phases are balanced in Eq.~(\ref{vlasovlinstand}) if the following conditions are satisfied
\begin{align}\label{fdistrbstand}
f_+&=-i \frac{q_s}{m_s} G_+\frac{1}{\omega-k v_x} \frac{\partial f_{0s}(v_{x})}{\partial v_{x}},\nonumber\\
f_-&=-i \frac{q_s}{m_s} G_-\frac{1}{\omega+k v_x} \frac{\partial f_{0s}(v_{x})}{\partial v_{x}}.
\end{align}
These expressions provide the fluctuation associated to each travelling wave and are in agreement with Eqs.~(\ref{eq:13}) and (\ref{eq:13left}) when we performed the calculations for each propagating wave independently. Hereafter we assume that $k$ is always positive. Now we can calculate the density fluctuation associated to the total distribution function based on Eq.~(\ref{totaldistrib}) by performing the integration in velocities. The result is 
\begin{align}\label{eq:15left1}
\delta n_{s}(x,t) &=i n_{0s}\frac{q_{s}}{m_{s}} \frac{G_+}{k w_{0s}^2} Z'(\xi_{s})\,e^{-i\omega t}e^{i k x}\nonumber \\
&-
i n_{0s}\frac{q_{s}}{m_{s}} \frac{G_-}{k w_{0s}^2} Z'(\xi_{s})\,e^{-i\omega t}e^{-ik x}.
\end{align}
Interestingly, this expression simplifies the problem if $G_+=-G_-$ since we obtain the simple form
\begin{equation}\label{npertplus}
\delta n_{s}(x,t) =i n_{0s}\frac{q_{s}}{m_{s}} \frac{G_+}{k w_{0s}^2} Z'(\xi_{s}) \,e^{-i\omega t}\left(e^{ik x}+e^{-ik x}\right).
\end{equation}
where we clearly have the superposition of the right and left propagating wave with the same amplitude. If we chose, that $G_+=G_-$ we simply have
\begin{equation}\label{npertminus}
\delta n_{s}(x,t) =i n_{0s}\frac{q_{s}}{m_{s}} \frac{G_+}{k w_{0s}^2} Z'(\xi_{s})\,e^{-i\omega t}\left(e^{ik x}-e^{-ik x}\right).
\end{equation}
Note that the addition of exponentials in Eq.~(\ref{npertplus}) reduces to a cosinus while the subtraction in Eq.~(\ref{npertminus}) leads to a sinus. This is precisely the profile we obtained previously, see for example Eq.~(\ref{solvpar2nm}).

\section{Forward and backward propagating waves}

It is straight forward to get from Eq.~(\ref{eq:11}) that
\begin{equation}\label{key1}
\frac{\partial f_{0s}(v_{x})}{\partial v_{x}}=\frac{n_{0s}}{\sqrt{\pi}w_{0s}}\frac{-2\,(v_{x}-u_{x0s})}{w_{0s}^2}\, e^{-\frac{(v_{x}-u_{x0s})^2}{w_{0s}^2}},
\end{equation}
\noindent therefore we have from Eq.~(\ref{eq:14}) that for a wave propagating to the right
\begin{equation}\label{key2}
\delta n_{s}(x,t) =i\frac{q_{s}}{m_{s}} \delta G_{s}(x,t) \frac{2\,n_{0s}}{\sqrt{\pi}w_{0s}^3} I_{s}(\omega , k),
\end{equation}
\noindent where
\begin{align}\label{key3}
I_{s}(\omega , k)=\int_{-\infty}^{+\infty} \frac{\left(v_{x}-u_{x0s}\right)\, e^{-\frac{(v_{x}-u_{x0s})^2}{w_{0s}^2}}}{\omega - k v_{x}} \,dv_{x},
\end{align}
by setting $\lambda=\frac{v_{x}-u_{x0s}}{w_{0s}}$ the integral $ I_{s}(\omega , k)$ becomes:
\begin{equation}\label{key4}
I_{s}(\omega , k)=-\frac{w_{0s}}{k} \int_{-\infty}^{+\infty} \frac{\lambda e^{-{\lambda}^2}}{\lambda - \xi_{s}}\, d\lambda,
\end{equation}
\noindent where $ \xi_{s}=\frac{\omega - k u_{x0s}}{k\, w_{0s}} $ which reduces to:
\begin{equation}\label{key5}
I_{s}(\omega , k) = \frac{\sqrt{\pi}w_{0s}}{2k}\,Z'(\xi_{s}),
\end{equation}
\noindent where we have defined $ Z'(\xi_{s})=-2\big(1+\xi_{s}\,Z(\xi_{s})\big) $ expressed in terms of the plasma dispersion function $ Z(\xi_{s}) $ (e.g. the Fried and Conte function) defined as
\begin{equation}\label{key6}
Z(\xi_{s}) = \frac{1}{\sqrt{\pi}} \int_{-\infty}^{+\infty} \frac{ e^{-{\lambda}^2}}{\lambda - \xi_{s}}\, d\lambda.
\end{equation}
Finally, substituting Eq.~(\ref{key5}) into Eq.~(\ref{key2}) we obtain Eq.~(\ref{eq:15}).  The integral form of the Zeta function is in principle only applicable when $Im (\xi_s) > 0$.

For the left propagating wave with wavenumber $-k$, we have
\begin{equation}\label{key3m}
I_{s}(\omega , -k)=\int_{-\infty}^{+\infty} \frac{v_{x}\, e^{-\frac{(v_{x})^2}{w_{0s}^2}}}{\omega + k  v_{x}} \,dv_{x},
\end{equation}

\noindent by setting $\lambda=-\frac{v_{x}}{w_{0s}}$ the integral now becomes:
\begin{align}\label{key4m}
I_{s}(\omega , -k )&=\frac{w_{0s}}{k} \int_{+\infty}^{-\infty} \frac{\lambda e^{-{\lambda}^2}}{\xi_{s}-\lambda }\, d\lambda=-\frac{w_{0s}}{k} \int_{-\infty}^{+\infty} \frac{\lambda e^{-{\lambda}^2}}{\xi_{s}-\lambda }\, d\lambda=\nonumber \\
& \frac{w_{0s}}{k} \int_{-\infty}^{+\infty} \frac{\lambda e^{-{\lambda}^2}}{\lambda -\xi_{s}}\, d\lambda
\end{align}
\noindent where $ \xi_{s}=\frac{\omega}{k w_{0s}} $, and the integral reduces to:
\begin{equation}\label{key5m}
I_{s}(\omega , -k) =- \frac{\sqrt{\pi}w_{0s}}{2 k}Z'(\xi_{s}).
\end{equation}
This expression is valid if $Im (\xi_s) > 0$. Note the difference in the sign between Eq.~(\ref{key5m}) (left propagating wave) and Eq.~(\ref{key5}) (right propagating wave). The previous analysis  takes into account the effect of the sign of $k$ in the dispersion relation, in agreement with the definition given in \citet{podestagary2011} (see their Eq.~(26)).

\section{The standing Alfv\'en wave solution}

Let us construct a standing Alfv\'enic wave solution using the results for circularly propagating waves. The idea is to superimpose two waves with the same frequency and same wavenumber but travelling in opposite directions. This is easily done in MHD because of the degeneracy of Alfv\'en waves. In the multifluid approach used in this work changing $\omega$ by $-\omega$ leads to another solution of the dispersion relation. Nevertheless, the symmetry regarding the wavenumber is still present, as we can see from Eq.~(\ref{disperalfm}). Therefore,  the propagation direction of the wave is easily reversed changing the sign of $k_0$.  We consider first right handed Alfv\'en waves. The perturbed velocities and magnetic fields components are of the following form
\begin{eqnarray}\label{eqvstandingf}
{u}_z&=&u_0 \cos{(k_0 x -\omega_0 t)},\\ 
B_z&=&-u_0 \frac{B_0}{v_{\rm A}^2}\frac{\omega_0}{k_0} \cos{(k_0 x -\omega_0 t)},\\
{u}_y&=&u_0 \sin{(k_0 x -\omega_0 t)},\\ 
B_y&=&-u_0 \frac{B_0}{v_{\rm A}^2}\frac{\omega_0}{k_0} \sin{(k_0 x -\omega_0 t)},
\end{eqnarray}
where we have used Eq.~(\ref{eqvperpbperpprot0}) to use the proper amplitude and sign of the magnetic field in terms of the initial velocity amplitude $u_0$. It is important to note that the amplitude of the perturbed magnetic field, $B_\perp=\sqrt{B_x^2+B_y^2}$ is a constant, independent of space and time (the same is also true for the perpendicular velocity).

The corresponding backward wave (we change $k_0$ by $-k_0$) has the same frequency (see Eq.~(\ref{disperalfm})) and (using again Eq.~(\ref{eqvperpbperpprot0})) we have
\begin{eqnarray}\label{eqvstandingg}
{u}_z&=&u_0 \cos{(-k_0 x -\omega_0 t)},\\ 
B_z&=&u_0 \frac{B_0}{v_{\rm A}^2}\frac{\omega_0}{k_0} \cos{(-k_0 x -\omega_0 t)},\\
{u}_y&=&u_0 \sin{(-k_0 x -\omega_0 t)},\\ 
B_y&=&u_0 \frac{B_0}{v_{\rm A}^2}\frac{\omega_0}{k_0} \sin{(-k_0 x -\omega_0 t)}.
\end{eqnarray}
Now we just need to make a linear combination of the two waves to obtain the standing solution. We choose to subtract the forward wave from the backward wave and using standard trigonometric formulas the result is written as
\begin{eqnarray}\label{eqvstandingr}
{u}_z&=&u_0 \sin(\omega_0 t) \sin(k_0 x) ,\\ 
B_z&=&-u_0 \frac{B_0}{v_{\rm A}^2}\frac{\omega_0}{k_0}  \cos(\omega_0 t) \cos{(k_0 x)},\\
{u}_y&=&u_0 \cos(\omega_0 t) \sin{(k_0 x)} ,\\ 
B_y&=&u_0 \frac{B_0}{v_{\rm A}^2}\frac{\omega_0}{k_0} \sin(\omega_0 t) \cos{(k_0 x)},
\end{eqnarray}
where a common multiplicative constant has been eliminated. It is evident from the previous expressions that, contrary to the propagating case, $B_\perp$ is not constant now. To properly excite a standing right hand wave in the simulations at $t=0$ we just need to impose a perturbation of the following form
\begin{eqnarray}\label{eqvstandingrl0}
{u}_z&=&0 ,\\ 
B_z&=&-u_0 \frac{B_0}{v_{\rm A}^2}\frac{\omega_0}{k_0}\cos{(k_0 x)},\\
{u}_y&=&u_0 \sin{(k_0 x)} ,\\ 
B_y&=&0.\label{eqvstandingrlf}
\end{eqnarray}

A similar approach is used to derive the expression for the standing left handed polarised wave. Now the dispersion relation is given by Eq.~(\ref{disperalfm}) and Eq.~(\ref{eqvperpbperpprot0}) does not change. For the forward wave we have
\begin{eqnarray}\label{eqvstandingrf}
{u}_z&=&-u_0 \cos{(k_0 x -\omega_0 t)},\\ 
B_z&=&u_0 \frac{B_0}{v_{\rm A}^2}\frac{\omega_0}{k_0} \cos{(k_0 x -\omega_0 t)},\\
{u}_y&=&u_0 \sin{(k_0 x -\omega_0 t)},\\ 
B_y&=&-u_0 \frac{B_0}{v_{\rm A}^2}\frac{\omega_0}{k_0} \sin{(k_0 x -\omega_0 t)},
\end{eqnarray}
while for the backward  wave ($k_0\rightarrow -k_0$) we have
\begin{eqnarray}\label{eqvstandingrb}
{u}_z&=&-u_0 \cos{(-k_0 x -\omega_0 t)},\\ 
B_z&=&-u_0 \frac{B_0}{v_{\rm A}^2}\frac{\omega_0}{k_0} \cos{(-k_0 x -\omega_0 t)},\\
{u}_y&=&u_0 \sin{(-k_0 x -\omega_0 t)},\\ 
B_y&=&u_0 \frac{B_0}{v_{\rm A}^2}\frac{\omega_0}{k_0} \sin{(-k_0 x -\omega_0 t)}.
\end{eqnarray}
Combining the previous waves we obtain
\begin{eqnarray}\label{eqvstandingl}
{u}_z&=&u_0 \sin(\omega_0 t) \sin(k_0 x) ,\\ 
B_z&=&-u_0 \frac{B_0}{v_{\rm A}^2}\frac{\omega_0}{k_0}  \cos(\omega_0 t) \cos{(k_0 x)},\\
{u}_y&=&-u_0 \cos(\omega_0 t) \sin{(k_0 x)} ,\\ 
B_y&=&-u_0 \frac{B_0}{v_{\rm A}^2}\frac{\omega_0}{k_0} \sin(\omega_0 t) \cos{(k_0 x)}.
\end{eqnarray}
Hence, to excite a circularly left handed Alfv\'en wave at $t=0$ we have to impose the following initial perturbation
\begin{eqnarray}\label{eqvstandingl0}
{u}_z&=&0 ,\\ 
B_z&=&-u_0 \frac{B_0}{v_{\rm A}^2}\frac{\omega_0}{k_0}\cos{(k_0 x)},\\
{u}_y&=&-u_0\sin{(k_0 x)} ,\\ 
B_y&=&0.\label{eqvstandinglf}
\end{eqnarray}

\section{Kinetic effects on ion acoustic waves driven by Alfv\'en waves}\label{sectkinetiondriven}

We carry out a new regular perturbation method on the Vlasov equation to introduce kinetic effects by implementing the method described in Section~\ref{kinetion}. To begin the new analysis scheme we first introduce the perturbation expansion of the macroscopic fluid quantities:
\begin{align}\label{expo2}
{\bf u}&= \epsilon\,{\bf u_{\perp}}+\epsilon^2\, {\delta u_{x}}\,{\bf {\hat e}_x}+\epsilon^3\,{\bf u'_{\perp}},\nonumber\\
{\bf B}&=B_0\,{\bf {\hat {e}}_x} + \epsilon\,{\bf B_{\perp}}+\epsilon^3\,{\bf B'_{\perp}},\nonumber\\
{\bf E}&=\epsilon\,{\bf E_{\perp}}+\epsilon^2 \delta E_x\, {\bf {\hat e}_x} +\epsilon^3\,{\bf E'_{\perp}} ,\nonumber\\
n&=n_0+\epsilon^2\, \delta n,\nonumber\\
p&=p_0+\epsilon^2\, \delta p.
\end{align}
The analysis begins with the linearisation scheme of the velocity distribution function
\begin{align}\label{expo2f}
F_{s}(x,{\bf v},t)&=F_{0s}({\bf v})+\epsilon^2\, \delta F_{s}(x,{\bf v},t).
\end{align}
Note that the perturbation expansion of the velocity distribution does not contain a first-order terms in $\epsilon$. This is because in this order an incompressible transverse Alfv\'en wave is imposed, which has no density perturbation associated with it. Furthermore, the zero-order expansion is chosen to be independent of time and space in the following form
\begin{align}\label{exp2}
F_{0s}({\bf v})&=\frac{n_{0s}}{\sqrt{\pi}w_{0s}} \delta ({\bf v}_\bot-{\bf{u}_{\perp}}_{0s})\, e^{-\frac{(v_{x}-u_{x0s})^2}{w_{0s}^2}}.
\end{align}
In these equations, again the velocity vector ${\bf v}$ corresponds to the particle distribution velocities whereas ${\bf{u}_{\perp}}_{0s}$ corresponds to the transverse Alfv\'en wave fluid velocity perturbation of the initial wave (given in Appendix C).

Inserting the perturbation expansion in Eqs.~(\ref{expo2}) and (\ref{expo2f}) into the Vlasov equation we obtain a hierarchy of equations at different orders in the perturbation expansion of the distribution function.
 At the zeroth order in $\epsilon$ the Vlasov equation is satisfied exactly for the equilibrium state. To first order in $\epsilon$, the Vlasov equation is also satisfied since there is no first order velocity distribution in the expansion nor first order density perturbation. At the second order in $\epsilon$ the Vlasov equation becomes
 
	\begin{align}\label{eq:9}
		\frac{\partial \delta F_{s}(x,{\bf v},t)}{\partial t}&+{\bf v}\cdot {\nabla}\delta F_{s}(x,{\bf v},t) \nonumber \\
		&+\frac{q_{s}}{m_{s}}\left({\bf E}(x,t) + \frac{{\bf v} \times {\bf B}(x,t)}{c}\right)^{(0)}_{x} \cdot\frac{\partial \delta F_{s}(x,{\bf v},t)}{\partial {\bf v}_{x}} \nonumber \\
		&+\frac{q_{s}}{m_{s}}\left({\bf E}(x,t) + \frac{{\bf v} \times {\bf B}(x,t)}{c}\right)^{(0)}_{\perp} \cdot\frac{\partial \delta F_{s}(x,{\bf v},t)}{\partial {\bf v}_{\perp}} \nonumber \\
		&+\frac{q_{s}}{m_{s}}\left({\bf E}(x,t) + \frac{{\bf v} \times {\bf B}(x,t)}{c}\right)^{(2)}_{x} \cdot\frac{\partial F_{0s}(x,{\bf v},t)}{\partial {\bf v}_{x}} \nonumber \\
		&+\frac{q_{s}}{m_{s}}\left({\bf E}(x,t) + \frac{{\bf v} \times {\bf B}(x,t)}{c}\right)^{(2)}_{\perp} \cdot\frac{\partial F_{0s}(x,{\bf v},t)}{\partial {\bf v}_{\perp}}=0.\nonumber \\
\end{align} 

\noindent where the force terms have been separated into parallel $(x)$ and perpendicular $(\perp) $ components relative to the mean magnetic field direction and the superscripts in parenthesis $(0)$ and $(2)$ represent zeroth and second order force terms defined below. Thus, to second order in $\epsilon$ the Vlasov equation reduces to
	\begin{align}\label{eq:9n}
		\frac{\partial \delta F_{s}(x,{\bf v},t)}{\partial t}&+{\bf v}\cdot {\nabla} \delta F_{s}(x,{\bf v},t) \nonumber \\
		=&-\frac{q_{s}}{m_{s}}\left(\delta E_{x}(x,t) + \frac{\left({\bf u_\perp}(x,t) \times {\bf B_\perp}(x,t)\right)_x}{c}\right) \cdot\frac{\partial F_{0s}({\bf v})}{\partial {\bf v}_{x}}.\nonumber\\
\end{align} 

\noindent At this point, in order to get a solution for the second order distribution $\delta F_{s}(x,{\bf v},t)$ we have to assume either a propagating or a standing solution. Note that Eq.~(\ref{eq:9n}) is equivalent to Eq.~(\ref{eq:12}) indicating that we it will eventually lead to the definition of the complex $\gamma$. 
\end{appendix}

\end{document}